\documentclass[journal=jacsat,manuscript=article]{achemso}
\setkeys{acs}{keywords = true}
\usepackage[version=3]{mhchem} 
\usepackage[table]{xcolor}
\usepackage{multirow}
\usepackage[normalem]{ulem}
\usepackage{hyperref}

\author{Hazem Abdelsalam}
\email{hazemabdelhameed@gmail.com}
\affiliation[]
{School of Materials Science and Engineering, Yancheng Institute of Technology, Yancheng 224051, P.R. China}
\alsoaffiliation[]
{Theoretical Physics Department, National Research Centre, Giza12622 Dokki, Egypt}
\author{Domenico Corona}
\affiliation[]
{Department of Physics, University of Rome Tor Vergata and INFN, Via della Ricerca Scientifica 1, 00133 Roma, Italy}
\email{domenico.corona@roma2.infn.it}
\affiliation[]
{Department of Physics, University of Rome Tor Vergata and INFN, Via della Ricerca Scientifica 1, 00133 Roma, Italy}
\author{Renebeth B. Payod}
\email{rbpayod@up.edu.ph}
\affiliation[]
{Institute of Physics, University of the Philippines, College, Los Ba\~{n}os, Laguna 4031 Philippines}

\author{Mahmoud A. S. Sakr}
\email{mahmoud.sakr@must.edu.eg}
\affiliation[]
{Chemistry Department, Center of Basic Science, Misr University for Science and Technology (MUST), P.O. 77, Giza, Egypt}
\author{Omar H. Abd-Elkader}
\email{omabdelkader7@ksu.edu.sa}
\affiliation[]
{Department of Physics and Astronomy, College of Science, King Saud University, P.O. Box 2455, Riyadh 11451, Saudi Arabia}

\author{Qinfang Zhang}
\email{qfangzhang@gmail.com}
\affiliation[]
{School of Materials Science and Engineering, Yancheng Institute of Technology, Yancheng 224051, P.R. China}

\author{Vasil A. Saroka}
\email{vasil.saroka@roma2.infn.it}
\affiliation[]
{Department of Physics, University of Rome Tor Vergata and INFN, Via della Ricerca Scientifica 1, 00133 Roma, Italy}
\alsoaffiliation[]
{TBpack Ltd., 27 Old Gloucester Street, London, WC1N 3AX, United Kingdom}
\alsoaffiliation[]
{Institute for Nuclear Problems, Belarusian State University, Bobruiskaya 11, 220030 Minsk, Belarus}

\title[]
  {Topological junction states in graphene nanoribbons: A route to topological chemistry}


\begin{document}

\begin{tocentry}




\centering
 \includegraphics[width=7cm]{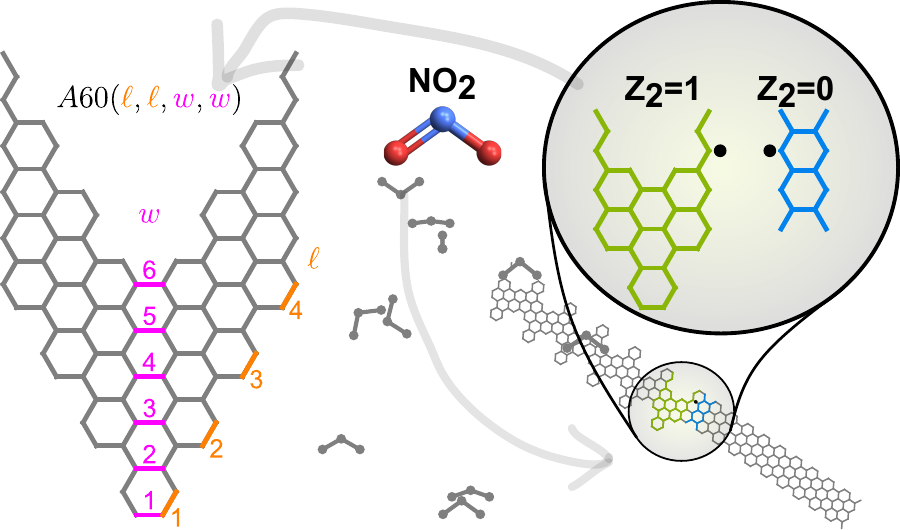}
\end{tocentry}

\begin{abstract}
Two-dimensional topological insulators with propagating topological edge states are promising for dissipationless transport, while their one-dimensional analogs are capable of hosting localized topological junction states that are mainly envisaged for quantum computing and spintronics. Here, in contrast, we propose to use the localized nature of topological junction states for sensing applications. We report a systematic topological classification of a wide class of graphene nanoribbons represented by already synthesized extended chevron species. Using this classification, we theoretically model a double junction transport that shows an enhanced interaction with the \ce{NO2} molecule. Our results show that topological junction states of nanoribbons can open an avenue for topological sensing and junction-assisted chemistry applications.
\end{abstract}

The exploration of topological states in condensed matter physics has revolutionized our understanding of electronic materials~\cite{Hasan2010,Ren2016,Cayssol2021}, opening new pathways for advanced technological applications in electronics and error-resilient quantum information~\cite{Weber2024}. Among the myriads of topologically intriguing systems~\cite{Qi2011a,Ando2013,Wehling2014a}, carbon-based nanostructures played a key role from the beginning~\cite{Kane2005a,Kane2005}. In recent years, particularly graphene nanoribbons (GNRs) have garnered significant attention~\cite{Cao2017,Lee2018,Lin2018,Ortiz2018,Rizzo2018,Groning2018,Joost2019,Jiang2021,Arnold2022,Kuo2023,Perkins2024,Wang2024,Zhao2024}. These one-dimensional (1D) nanostructures, derived from graphene, exhibit unique electronic properties due to their distinct edge terminations and narrow widths~\cite{Yano2020,Chen2020}. 

The bulk-boundary correspondence principle is at the heart of the topological band theory of solids. It states that, at the interfaces between regions of different topological orders, topological interface states occur as highly localized electronic modes that are resilient to disorder or perturbations~\cite{Hasan2010}. The topological stability of these interface states can be controlled by several topological invariants. In the case of the integer quantum Hall effect~\cite{Simon1983,Niu1985} or Haldane model~\cite{Haldane1988}, the invariant is the Chern number, while it is $Z_2$ invariant instead for topological insulators~\cite{Kane2005a,Kane2005}. The $Z_2$ invariant provides the binary classification of topological phases with values of $0$ or $1$. A wider classification based on the Chern number utilizes the whole set of integer numbers $Z$. In both cases, a zero value stands for a trivial topological order. An interface state appears, whenever two regions with different $Z$ or $Z_2$ values meet each other. This basic principle works in all dimensions and has recently attracted a lot of attention for 1D structures, such as graphene nanoribbons, where topological junction states (TJS) have been revealed and classified~\cite{Cao2017,Lee2018,Jiang2021,Arnold2022}. The classification of TJS in GNRs can be carried out based on $Z_2$ invariant~\cite{Cao2017,Lin2018,Lee2018} and also can be generalized by incorporating chiral symmetry and spin corrections~\cite{Jiang2021}. Some of these states have already been experimentally shown and investigated through an atomically precise on-surface bottom-up engineering~\cite{Wang2016a,Rizzo2018,Groning2018,Wang2024}.

The envisaged applications of topologically protected states are (i) disorder-resilient low-dissipation transport and (ii) robust long-coherence spin states~\cite{Ren2016,Weber2024}. The latter  has become a roadmap for topological junction states as potential candidates for designing customizable spintronics~\cite{Cao2017,Ortiz2018,Perkins2024} and quantum computing~\cite{Ortiz2018,Perkins2024} nanodevices. Notably, this widely accepted roadmap overlooks chemical sensing, which is arguably a more natural and immediate application. Chemical sensing relies critically on the interaction between the sensor material and the target analyte, where sensitivity and selectivity are paramount~\cite{Gao2016}. The 1D nature of GNRs, combined with topological junctions providing a high density of localized electronic states to facilitate interaction with chemical species, potentially offers enhanced sensitivity and selectivity that are accessible for readout in a standard transport device similar to a field-effect transistor~\cite{Adamu2021,Johnson2021}. 

This paper presents a theoretical investigation of topological junction states in GNRs, with a particular focus on their application in gas sensing. The gas sensing application has already been proposed and investigated both theoretically~\cite{Abdelsalam2019b,Cernevics2020,Abdelsalam2021,Sakr2024} and experimentally~\cite{MehdiPour2017,Cho2018} for the edge and end states of GNRs; see also a review of more general sensing schemes in Ref.~\citenum{Johnson2021}. Some experimental studies have demonstrated the detection of volatile compounds such as ethanol and methanol~\cite{MehdiPour2017} as well as nitrogen dioxide \ce{NO2}~\cite{Cho2018}. However, the sensitivity of TJS in GNRs remains unknown. Here, we explore the electronic properties of these states, their formation mechanisms, and their interaction with \ce{NO_2} as the target molecule. We have noticed that some of the ribbons that we have studied previously~\cite{Abdelsalam2021} (we refer to those as A60 within a more general class of zigzag-shaped GNRs~\cite{Saroka2014a,Saroka2015,Saroka2015a,Saroka2016a}), are, in fact, topologically nontrivial. For some combinations of their unit cell structural parameters, they are characterized by the $Z_2$ invariant derived from the intercellular Zak phase~\cite{Kudin2007,Rhim2017} calculated in the periodic gauge of the tight-binding (TB) Hamiltonian~\cite{Cayssol2021}. Thus, they can form TJSs via seamless combination with a topologically trivial unit cells of armchair GNRs (AGNRs)~\cite{Cao2017}. Performing density functional theory (DFT) calculations to study sensing, we focus on nitrogen dioxide (\ce{NO2}) gas, which is a free-radical and a moderate Lewis acid. \ce{NO2} is an air pollutant affecting the environment and health of people~\cite{Last1994,Huangfu2020} and a prototype of a vast range of substances belonging to nitrocompound group, e.g. nitroparaffins and nitroarenes.
By leveraging the unique properties of topological junctions in GNRs, we aim to elucidate their potential in creating highly sensitive and selective chemical sensors, paving the way for advancements in environmental monitoring, and industrial safety~\cite{Li2022}.

{\bf Topological properties.} 
The TB Hamiltonian of $\pi$-electron network is (neglecting a vanishing spin-orbit term~\cite{Sichau2019})
\begin{equation}
  H= \sum_i \varepsilon c^{\dagger}_i c_i + \sum_{i,j} t_{1} \left( c^{\dagger}_i c_j + c^{\dagger}_j c_i\right)
  \label{eqn:TBHamiltonian}
\end{equation}
where $c^{\dagger}_i$ and $c_i$ are electron creation and annihilation operators, respectively, $\varepsilon=0$ is the on-site energy equal for all lattice sites, $t_{1}=3.12$~eV~\cite{Partoens2006} is the nearest-neighbor hopping integral. We note here that $t_1$ varies for different structures~\cite{Saroka2018a,Payod2020}, therefore, we will use $E/t_1$ dimensionless units in what follows. By going into reciprocal $k$-space in Eq.~(\ref{eqn:TBHamiltonian}) via a Fourier transform, $c^{\dagger}_i = \tfrac{1}{\sqrt{N}} \sum_{k} e^{-\mathrm{i} k R_i} c^{\dagger}_{k}$ (and similar for $c_i$) with $N$ being the number of unit cells and $R_i$ being the site position, we solve the eigenproblem of $n \times n$ matrix Hamiltonian for a given unit cell of a GNR with $n$ lattice sites in the cell, using \emph{TBpack} package~\cite{TBpackSaroka,SM-nl-2025-001492}. Then $k$-space Hamiltonian eigenvectors, $C_{i}(k) = \left(C_{i,1}(k), C_{i,2}(k),\ldots, C_{i,n}(k)\right)^{\mathrm{T}}$, can be employed for the calculation of the intercellular Zak phase by Eq.~(31) in Ref.~\citenum{Kudin2007}:
\begin{equation}
  \gamma_2= -\mathrm{Im} \left\{\log \prod_{j=1}^{N_k} \mathrm{Det} \left[C^{\dagger}_p (k_{j}) C_q (k_{j+1})\right]\right\}\, ,
  \label{eqn:InterZakPhase}
\end{equation}
where $N_k$ is the number of $k$-point sampling the Brillouin zone of the GNR, and the indices $p$ and $q$ run from $1$ to $n/2$ so that dot products of eigenvectors in Eq.~(\ref{eqn:InterZakPhase}) form a matrix from all occupied states.

\begin{figure}
\includegraphics[width=0.95\textwidth]{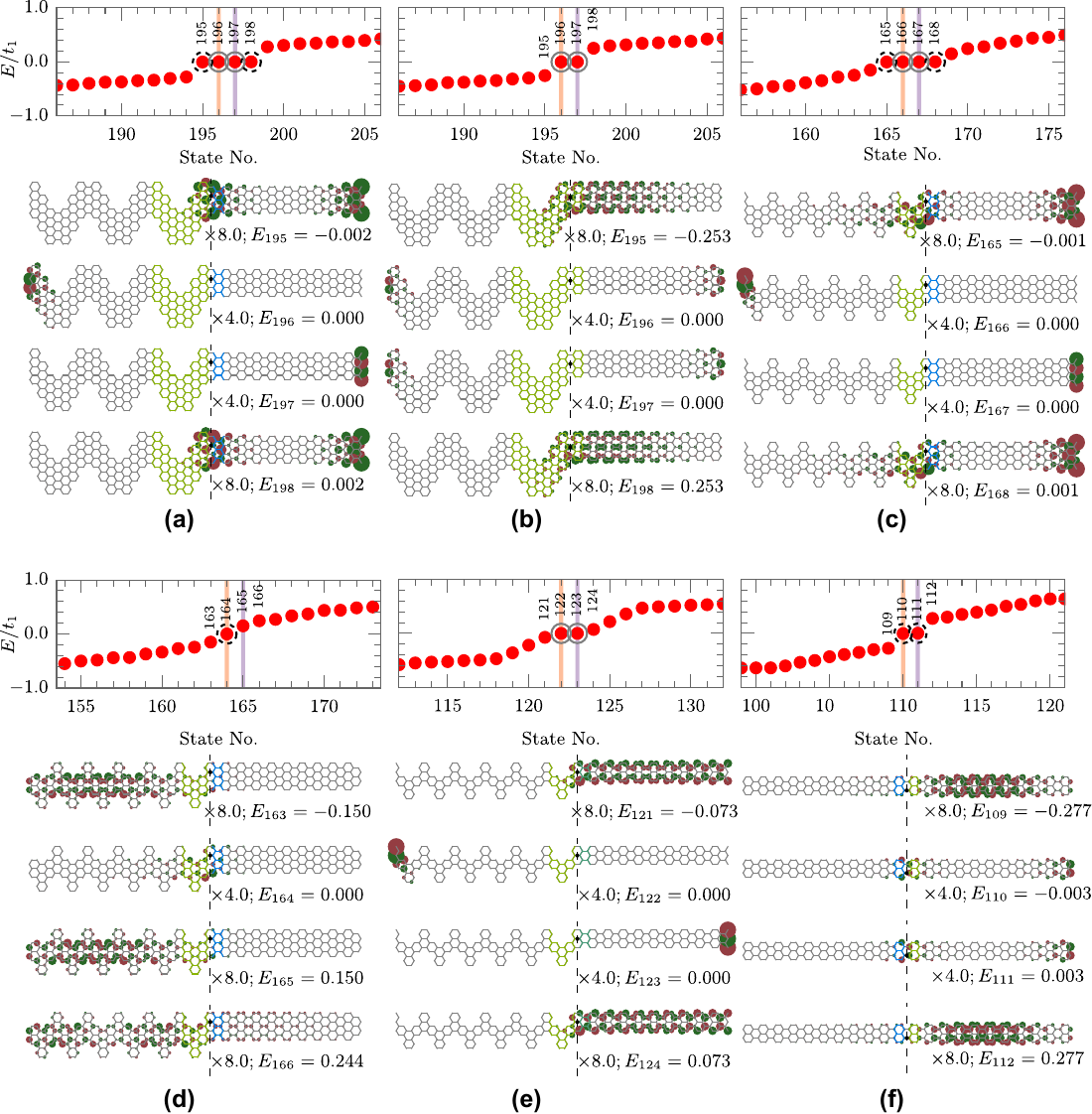}
  \caption{Electronic properties of topological junctions. (a) Energy levels for cGNR-AGNR($7$) topological-trivial junction. Vertical orange and violet lines mark $\lceil N/2\rceil$ and $\lceil N/2\rceil+1$, with $\lceil x \rceil$ and $N$ being the ceiling operation and total number of C atoms in the structure, respectively. Dashed black circles highlight TJSs, gray circles highlight end states. The wave functions of the numbered states are presented below, where the scaling factors together with the state energies in terms of $E/t_1$ are given. Dark red and green show the positive and negative phase of the wave function, respectively. Vertical dashed black lines denote the junction. Black dot is the alignment point for the two underlined unit cells. Light green and light blue highlight topological and trivial unit cells, respectively. (b) Same as (a) but for cGNR-AGNR($7$) topological-topological junction. (c) Same as (a) but for A60$(2,2,4,4)$-AGNR($7$) topological-trivial junction. (d) Same as (a) but for A60$(2,2,4,4)$-AGNR($7$) junction with modified ends. (e) Same as (a) but for A60$(2,2,3,3)$-AGNR($5$) ill-defined junction. The green blue color highlights the ``metallic" unit cell. (f) Same as (a) but for the edge-strain-induced junction in AGNR($5$): $t_{\mathrm{edge}}=0.5 t_1$ to the left  and $1.5 t_1$ to the right of the junction.}
  \label{fig:TopologicalJunctionStatesTB}
\end{figure}

In Table~\ref{fig:TopologicalJunctionStatesTB}, six topological junctions of interest are constructed. 
\begin{table}
  \caption{Junctions between various types of GNRs. Light green marks topological structures, while light blue and green-blue - trivial and undefined, i.e. ``metallic", structures, respectively. Black dot is the alignment point for a pair of unit cells forming a junction. Crimson dots show excluded from $\pi$-network C atoms that are not bonded to as many C atoms as they would be in the bulk of the structure. Insets: possible chemical terminations. The period of translation for the bulk structures is along horizontal $Ox$-axis.}
  \label{tbl:TypesOfGNRJunctions}
 \begin{tabular}{c|c|c|c}
   \hline
  Case & Unit cell 1 & Unit cell 2 & Junction \\
   \hline
   \multirow{3}*{1\textsuperscript{\emph{a}}} & cGNR & AGNR($7$) & cGNR-AGNR($7$)\\
    & $Z_2 = 1$ & $Z_2=0$ & \\ 
    & \includegraphics[width=1.8cm]{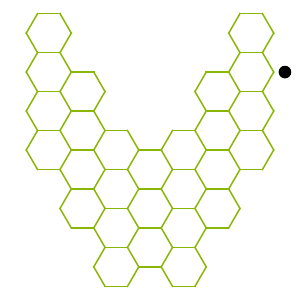} & \includegraphics[width=1.8cm]{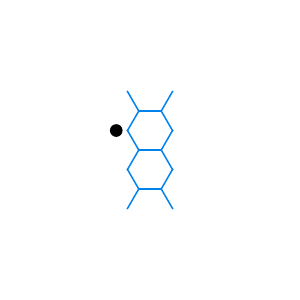} & \includegraphics[width=9cm]{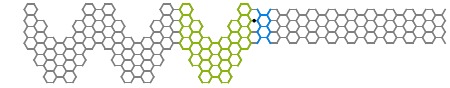} \\ \hline
\multirow{3}*{2\textsuperscript{\emph{a}}} & cGNR & AGNR($7$) & cGNR-AGNR($7$)\\
    & $Z_2 = 1$ & $Z_2=1$ & \\ 
    & \includegraphics[width=1.8cm]{Figs/Table1/cGNR_uc.pdf} & \includegraphics[width=1.8cm]{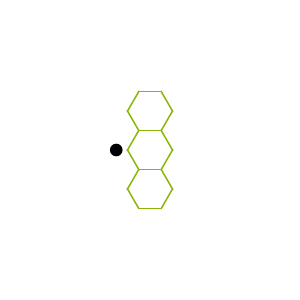} & \includegraphics[width=9cm]{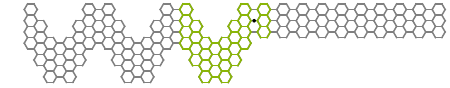} \\ \hline
    \multirow{3}*{3} & A60$(2,2,4,4)$ & AGNR($7$) & A60$(2,2,4,4)$-AGNR($7$)\\
    & $Z_2 = 1$ & $Z_2=0$ & \\ 
    & \includegraphics[width=1.9cm]{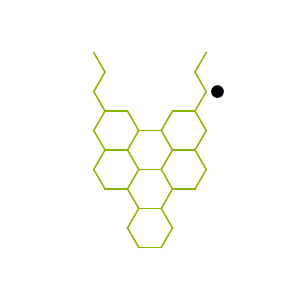} & \includegraphics[width=1.9cm]{Figs/Table1/AGNR_7_uc_typeII.pdf} & \includegraphics[width=9cm]{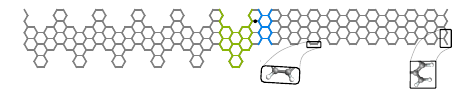} \\ \hline
    \multirow{3}*{4} & A60$(2,2,4,4)$ & AGNR($7$) & A60$(2,2,4,4)$-AGNR($7$) modified ends\\
    & $Z_2 = 1$ & $Z_2=0$ & \\ 
    & \includegraphics[width=1.8cm]{Figs/Table1/A60-2,2,4,4_uc.pdf} & \includegraphics[width=1.8cm]{Figs/Table1/AGNR_7_uc_typeII.pdf} & \includegraphics[width=9cm]{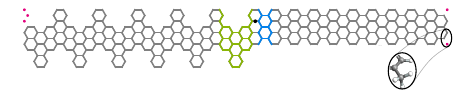} \\ \hline
     \multirow{3}*{5} & A60$(2,2,3,3)$ & AGNR($5$) & A60$(2,2,3,3)$-AGNR($5$)\\
    & $Z_2 = 1$ & $Z_2=\{0,1\}$\textsuperscript{\emph{b}} & \\ 
    & \includegraphics[width=1.8cm]{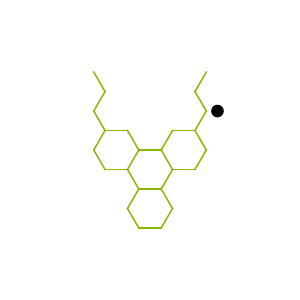} & \includegraphics[width=1.8cm]{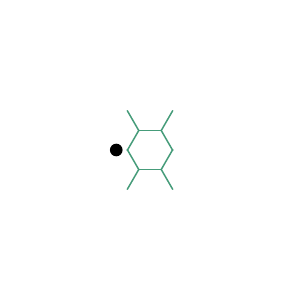} & \includegraphics[width=9cm]{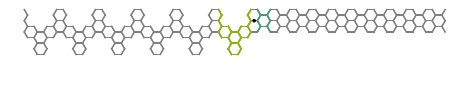} \\
\hline
    \multirow{3}*{6\textsuperscript{\emph{c}}} & AGNR($5$) & AGNR($5$) & AGNR($5$)-AGNR($5$)\\
    & $Z_2 = 0; t_{\mathrm{edge}}<t_1$ & $Z_2=1; t_{\mathrm{edge}}>t_1$ & \\ 
    & \includegraphics[width=1.8cm]{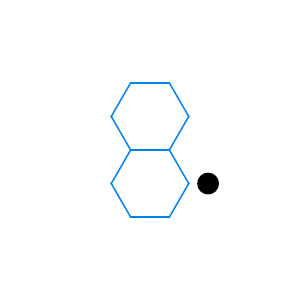} & \includegraphics[width=1.8cm]{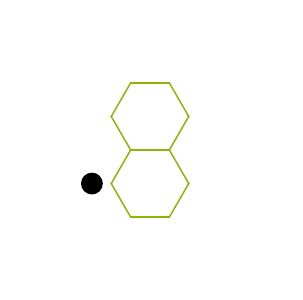} & \includegraphics[width=9cm]{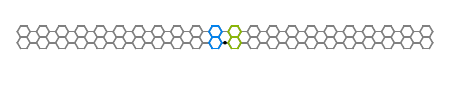} \\
\hline
  \end{tabular}

 \textsuperscript{\emph{a}} Junctions previously reported, see Fig.~3(b) in Ref.~\citenum{Lee2018};
 \textsuperscript{\emph{b}} Ambiguity of $Z_2$ with respect to applied strain has been reported in Ref.~\citenum{Tepliakov2023}; see, for example, Fig.~3(c); Here the structure is considered to be gapless, i.e. metallic one;
 \textsuperscript{\emph{c}} This junction is obtained by reducing model of Ref.~\citenum{Tepliakov2023}, that accounts  up to third nearest neighbor hopping interactions.
\end{table}
The first two cases are dealing with chevron-type GNR (cGNR)~\cite{Cai2010} and armchair GNRs (AGNRs), classified by the number of carbon atom pairs in their unit cells $N_p$, i.e. AGNR($N_p$). Those Cases~1 and~2 are used to verify topological band theory in finite-size cluster approach to supplement periodic calculations of Refs.~\citenum{Cao2017,Lee2018}. Cases~3 and~4 present the topological junctions in focus of the given study, constructed using the A60 class of GNRs~\cite{Saroka2014a,Saroka2015,Saroka2016a} defined by a set of four structural parameters $(\ell_1, \ell_2; w_1, w_2)$, that are explained in Scheme~\ref{sch:A60symStructure}. 
\begin{scheme}
  \includegraphics{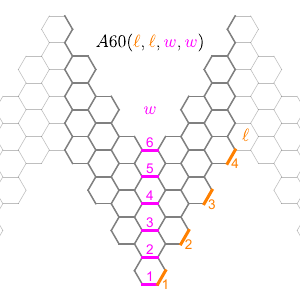}
  \caption{Main structural parameters of a unit cell of a mirror symmetric A60 class of GNRs: $\ell_1=\ell_2=\ell$ and $w_1=w_2=w$. A60$(\ell_1,\ell_2,w_1,w_2)\Longrightarrow$ A60$(4,4,6,6)$ is taken as an example. It was synthesized in Ref.~\citenum{MehdiPour2017}. The period of translation is along horizontal axis.}
  \label{sch:A60symStructure}
\end{scheme}
The $Z_2$ topological classification of symmetric A60 GNRs, $\ell=\ell_1=\ell_2$ and $w = w_1 = w_2$, is summarized in Table~\ref{tbl:Z2indexForA60}. For the sake of completeness, Cases~5 and~6 dealing with metallic structures~\cite{Son2006a} and strain~\cite{Tepliakov2023} are added to Table~\ref{fig:TopologicalJunctionStatesTB}.
\begin{table}
  \caption{$Z_2$ topological invariant of A60$(\ell_1, \ell_2; w_1, w_2)$ GNRs as a function of their structural parameters $\ell=\ell_1=\ell_2$ and $w=w_1=w_2$. Ligthgray: A60 GNRs used further in this study. Footnotes: species previously studied in some  aspects.}
  \label{tbl:Z2indexForA60}
 \begin{tabular}{c|cccccccccccccccccc}
   \hline
   $\ell$ \textbackslash $w$ & 3 & 4 & 5 & 6 & 7 & 8 & 9 & 10 & 11 & 12 & 13 & 14 & 15 & 16 & 17 & 18 & 19 & 20 \\
   \hline
2 & \cellcolor{lightgray} 1 & \cellcolor{lightgray}1 & 0 & 0 & 0 & 1 & 1 & 1 & 0 & 0 & 0 & 1 & 1 & 1  & 0 & 0 & 0 & 1 \\
3 & 1\textsuperscript{\emph{c}}& 1 & 0 & 0\textsuperscript{\emph{d}} & 0 & 1 & 1 & 1 & 0 & 0 & 0 & 1 & 1 & 1 & 0 & 0 & 0 & 1 \\
4 & 1 & 1\textsuperscript{\emph{a}} & 0 & 0\textsuperscript{\emph{b}} & 0 & 1 & 1 & 1 & 0 & 0 & 0 & 1 & 1 & 1 & 0 & 0 & 0 & 1 \\
5 & 1 & 1 & 0 & 0 & 0 & 1 & 1 & 1 & 0 & 0 & 0 & 1 & 1 & 1 & 0 & 0 & 0 & 1 \\
6 & 0 & 1 & 0 & 0 & 0 & 1 & 0 & 1 & 0 & 0 & 0 & 1 & 0 & 1 & 0 & 0 & 0 & 1 \\
7 & 0 & 1 & 0 & 0 & 0 & 1 & 1 & 1 & 0 & 0 & 0 & 0 & 1 & 1 & 0 & 0 & 0 & 1 \\
8 & 0 & 1 & 0 & 0 & 0 & 0 & 0 & 0 & 0 & 0 & 0 & 0 & 0 & 0 & 0 & 0 & 0 & 0 \\
9 & 0 & 1 & 0 & 0 & 0 & 1 & 0 & 1 & 0 & 0 & 0 & 0 & 1 & 0 & 0 & 0 & 0 & 0 \\
10 & 0 & 1 & 0 & 0 & 0 & 0 & 0 & 0 & 0 & 0 & 0 & 0 & 0 & 0 & 0 & 0 & 0 & 0 \\
\hline
\end{tabular}

 \textsuperscript{\emph{a}} Fig.~2 in Ref.~\citenum{Saroka2014a};
 \textsuperscript{\emph{b}} Fig.~2(b) in Ref.~\citenum{Lee2018} and Fig.~1 in Ref.~\citenum{MehdiPour2017};
  \textsuperscript{\emph{c}} Fig.~2(a,d) in Ref.~\citenum{Abdelsalam2021} ;
  \textsuperscript{\emph{d}} Fig.~2(b,e) in Ref.~\citenum{Abdelsalam2021}.
\end{table}

Figure~\ref{fig:TopologicalJunctionStatesTB} summarizes the results for the nanoribbon junctions presented in Table~\ref{tbl:TypesOfGNRJunctions} and the states observed for them. 
As one can see from Figure~\ref{fig:TopologicalJunctionStatesTB}a, describing the finite-size structure of Case~1 in Table~\ref{tbl:TypesOfGNRJunctions}, the zero energy hosts four states, two of which, i.e. No.~$195$ and~$198$, are localized at the junction region marked with a vertical dashed black line. These states also have some weight at the trivial end of the structure. This feature is missing in periodic structure calculations and never mentioned in Ref.~\citenum{Lee2018}. The two other states, No.~$196$ and No.~$197$, are the end states. The state No.~$196$ is a topological end state localized at the border of the topological cGNR unit cell and the vacuum, which is trivial insulator, $Z_{2} =0$, because it can be explained by the bulk-boundary correspondence. In contrast, the state No.~$197$ forms at the interface between trivial AGNR($7$) unit cell (light blue color) and the trivial vacuum which must be an ordinary end state related to carbon atoms featuring unpaired electrons as we shall see later. Figure~\ref{fig:TopologicalJunctionStatesTB}b shows the energy levels and the wave functions of the cGNR-AGNR($7$) junction, which is Case~2 in Table~\ref{tbl:TypesOfGNRJunctions}. In this structure, there are only two zero-energy states, which are end states. These states No.~$196$ and~$197$, respectively, are both localized at the interface between a topological unit cell and a trivial vacuum. Thus, these states are predicted by the bulk-boundary correspondence principle so that they are both topological end states. For comparison, the two bulk states are also presented in Figure~\ref{fig:TopologicalJunctionStatesTB}b.
Analogous relations can be noticed in Figure~\ref{fig:TopologicalJunctionStatesTB}c corresponding to Case~3 in Table~\ref{tbl:TypesOfGNRJunctions}. Here we show electronic energy levels and wave functions for A60$(2,2,4,4)$-AGNR($7$) junction between topological and trivial unit cells, respectively. Similar to Figure~\ref{fig:TopologicalJunctionStatesTB}a, along with the two TJSs, No.~$165$ and~$168$, having some weight at the end of the structure, there are states No.~$166$ and~$167$ localized entirely at the ends of the structure. To verify the hypothesis that they are merely related to the C atoms featuring unpaired electrons, we constructed Case~4 structure with $5$~atoms removed as shown in Table~\ref{tbl:TypesOfGNRJunctions}. Figure~\ref{fig:TopologicalJunctionStatesTB}d confirms that only one TJS remains after all the C atoms carrying unpaired electrons have been removed. Finally, we present the results for the controversial Case~5. We note that Eq.~(\ref{eqn:InterZakPhase}) applied to the AGNR($5$) yielded $Z_2 = 1$. Despite the appeal to interpret the results in Figure~\ref{fig:TopologicalJunctionStatesTB}e with the bulk-boundary correspondence as in the Case~2 discussed with Figure~\ref{fig:TopologicalJunctionStatesTB}b, this approach will not work for other ``metallic" structures. Due to this reason, the unit cell of AGNR($5$) is highlighted with a blend between light green and light blue colors. On the other hand, when the metallic structure has a well-defined gap due to the edge strain as in Case~6, then a localized TJS is clearly identifiable, as one can see from Figure~\ref{fig:TopologicalJunctionStatesTB}(f). Note that to construct the strain junction the unit cell must be chosen differently; see Ref.~\citenum{Cao2017} Table~1 for allowed choices. Thus, we have seen that topological band theory allows us to predict and reliably construct localized junction states. From a chemistry point of view, these nucleophilic regions with excess electrons due to TJSs are similar to radicals and they must be perfect reaction sites. The geometry of the junction compatible with a standard transport experiment and radical-like nature of the TJS must make them useful for sensing application. 

\begin{figure}
  \includegraphics[width=0.8\textwidth]{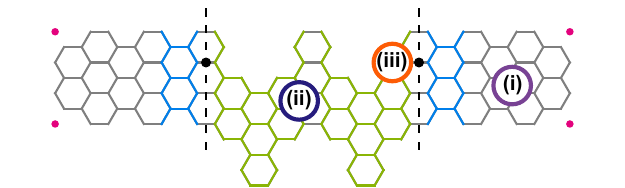}
  \caption{Double junction scheme. Following notation used in Table~\ref{tbl:TypesOfGNRJunctions}, the light green unit cells of A60$(2,2,4,4)$ are topological ones, while light blue unit cells of AGNR($7$) are trivial. The alignment points for the two junctions are depicted with black dots, while the junctions are marked with two dashed black lines. The C atoms that are excluded from $\pi$-electron network by their conversion into methyl groups are shown as crimson dots. The three circles highlight the chosen site for investigating adsorption properties: (i) violet - AGNR($7$), (ii) dark blue - A60$(2,2,4,4)$, and (iii) orange - one of the two topological junctions. Hydrogen atoms are not displayed.}
  \label{fig:DJDStructure}
\end{figure}

{\bf Topological sensing.} The above presented junctions are all asymmetric. The asymmetry of the junctions may pose a significant problem for potential transport measurements by requiring differing fabrication approaches of the left and right leads. This problem could be bypassed in a mirror symmetric double junction (DJ). As shown in Figure~\ref{fig:DJDStructure}, a DJ has a few topological unit cells of an A60 GNR sandwiched between trivial AGNRs.

Using DFT calculations as implemented in Gaussian~16~\cite{g16} and postprocessing with \emph{Multiwfn} software~\cite{Lu2012} for partial density of states (PDOS) extraction, we scrutinized the electronic and adsorption properties of the DJ~\cite{SM-nl-2025-001492}. See further details in the Supplementary Information Note~1.
In the constructed double junction, the C atoms with unpaired electrons are electronically saturated via hydrogenation to form methyl groups, therefore, we expect one TJS per junction similar to Figure~\ref{fig:TopologicalJunctionStatesTB}d. Since two junctions are placed close to each other in our DJ, the two TJS can slightly overlap. This interaction between TJSs shall result in their symmetric splitting around the Fermi energy, $E_{\mathrm{f}} = 0$~eV, similar to what happens with the ground state of a double quantum well~\cite{Collier2017a}.

\begin{figure}
\includegraphics[width=0.85\textwidth]{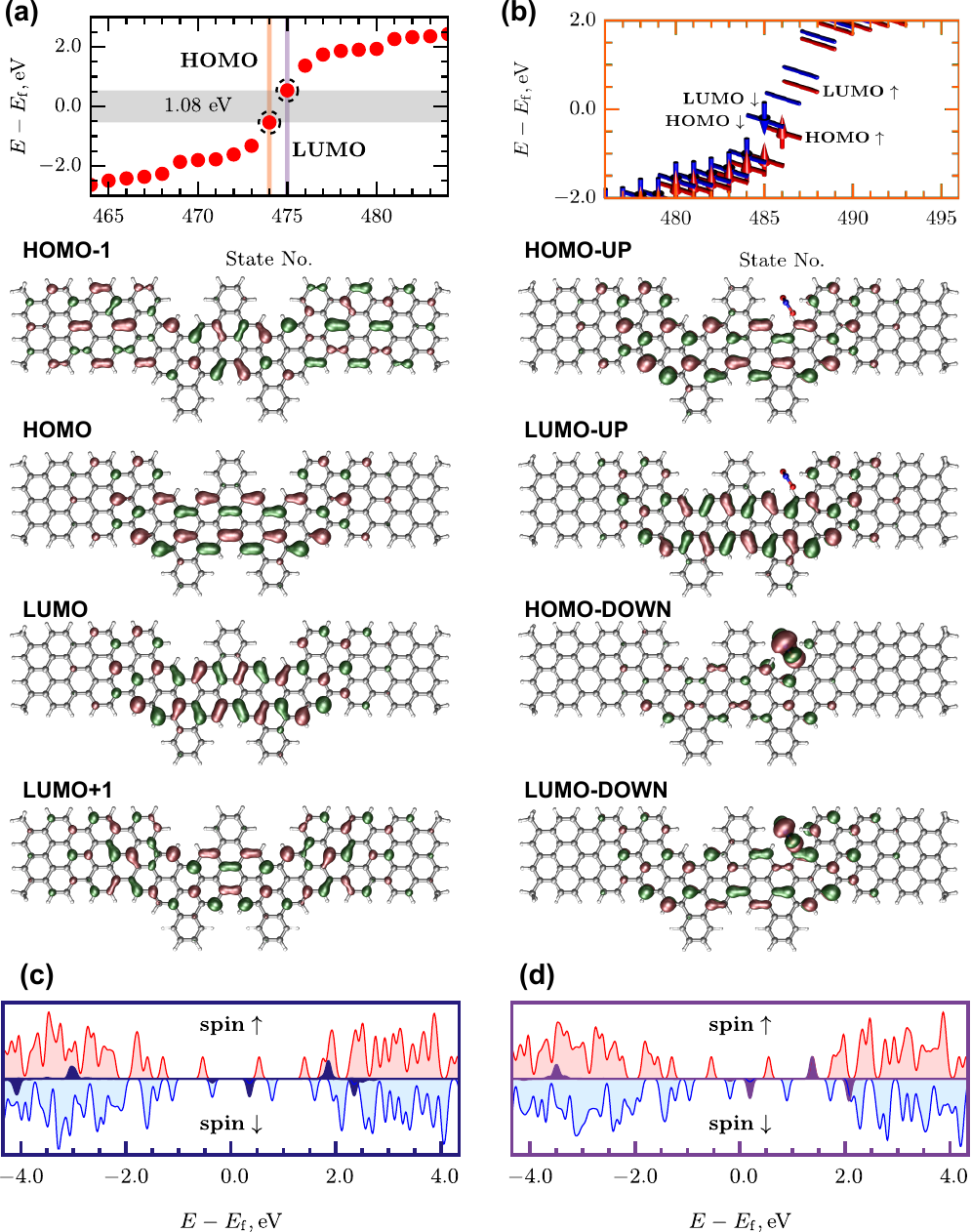}
  \caption{Interaction with \ce{NO2} gas molecule. (a) The electronic energy levels of pristine DJ together with molecular orbitals of TJS and extended bulk states. Isovalue: $0.02$. Similar to Figure~\ref{fig:TopologicalJunctionStatesTB} vertical orange and violet lines mark HOMO and LUMO, while dashed black circles highlight TJSs. (b) The spin-polarized electronic energy levels for DJ after adsorption of \ce{NO2} molecule at the A60$(2,2,4,4)$-AGNR($7$) junction between topological and trivial unit cells. (c) The spin-resolved partial density of states for DJ and \ce{NO2} molecule adsorbed at A60$(2,2,4,4)$. The \ce{NO2} data are plotted by the same dark blue color that is used to denote \ce{NO2} adsoption site in Figure~\ref{fig:DJDStructure}. (d) Same as (c), but for \ce{NO2} adsorbed at AGNR($7$). Broadening: $0.1$~eV. The frame colors in (b), (c), and (d) correspond to those of circles showing the adsoption sites in Figure~\ref{fig:DJDStructure}.}
  \label{fig:NO2adsorption}
\end{figure}

As one can see from Figure~\ref{fig:NO2adsorption}(a), the DFT modeling confirms the expected ground state splitting. The DJ highest occupied molecular orbital (HOMO) and lowest unoccupied molecular orbital (LUMO) energies are separated by the gap $E_{\mathrm{g}} = 1.08$~eV. The distributions of the HOMO and LUMO are clearly different from those of the bulk orbitals. While bulk states extend over a large number of trivial AGNR($7$) unit cells, the HOMO and LUMO representing TJSs are localized at the two A60$(2,2,4,4)$ unit cells in the region between the two junctions. These topological states are expected to be more interactive than the bulk ones. To compare TJSs to other energy states in the DJ, the reactivity is tested by studying the adsorption properties of the \ce{NO2} gas molecule on different sites of the DJ. In this testing, the \ce{NO2} is initially positioned at about $3$~{\AA} above the DJ at three sites:  (i) the AGNR, (ii) the A60 GNR, and (iii) the junction. Then the whole structure is optimized so that \ce{NO2} ends up at the sites illustrated in Figure~\ref{fig:DJDStructure}. The three chosen adsorption configurations cover reduced electron densities away from the junction and increased electron density due to TJS in the junction region. The adsorption energies in the first two cases are comparable, namely $-0.27$ and $-0.30$~eV, respectively. In contrast, the adsorption energy at the junction site is greater than double of these values, i.e. $E_{\mathrm{a}}=-0.66$~eV. This indicates that TJSs indeed boost interaction with the target molecule.  
Adsorption of \ce{NO2} is accompanied by the electron charge transfer of $-0.55e$ Mulliken charge [$-0.43e$ Hirshfeld charge] from the junction to \ce{NO2} molecule. Before the adsorption, the O-atom charge in \ce{NO2} molecule is $-0.20e$ [$-0.1035e$] for both oxygen atoms, while N-atom charge is $0.40e$ [$0.207e$]. However, these charges decrease to $-0.42e$ [$-0.247e$] and $-0.35e$ [$-0.236e$] for oxygen atoms and to $0.22e$ [$0.053e$] for the nitrogen atom after adsorption at the junction site. 
For comparison, adsorption of \ce{NO2} on the A60 and AGNR sites results in $-0.19e$ [$-0.155e$] and $-0.11e$ [$-0.081e$] Mulliken [Hirshfeld] charge transfers, respectively. Therefore, physical adsorption at the junction site is stronger than on the other two sites. Figure~\ref{fig:NO2adsorption}(b) shows spin-polarized energy levels for the system, wherein \ce{NO2} is adsorbed on the junction spot of the DJ. The main effect of the \ce{NO2} adsorption is induced spin-splitting of the TJSs, both HOMO and LUMO, and the resulting band gap reduction for spin-down component: $E_{\mathrm{g},\downarrow} = 0.51$~eV, as compared to $E_{\mathrm{g},\uparrow} = 1.04$~eV and $E_{\mathrm{g}} = 1.08$~eV of a spin-unpolarized case in Figure~\ref{fig:NO2adsorption}(a). As seen from the HOMO spin-down component, the contribution of \ce{NO2} into the spin-down component of the bonding HOMO is significant in the case of adsorption at the junction which is compelling evidence of the localized TJSs being more reactive than extended bulk states. The partial densities of states plotted in Figure~\ref{fig:NO2adsorption}(c) and (d) further
demonstrate that \ce{NO2} contributions (dark blue and violet) into HOMOs are almost vanishing when the adsorption takes place at A60$(2,2,4,4)$ and AGNR($7$) sites even though the energy gaps are comparable to the junction case: $E_{\mathrm{g},\downarrow} = 0.71$~eV and $0.38$~eV and 
$E_{\mathrm{g},\uparrow} = 1.07$~eV and $1.09$~eV, respectively. 
This indicates that TJSs may be suitable for facilitating sensing and catalytic applications.

\begin{figure}
\includegraphics[width=0.8\textwidth]{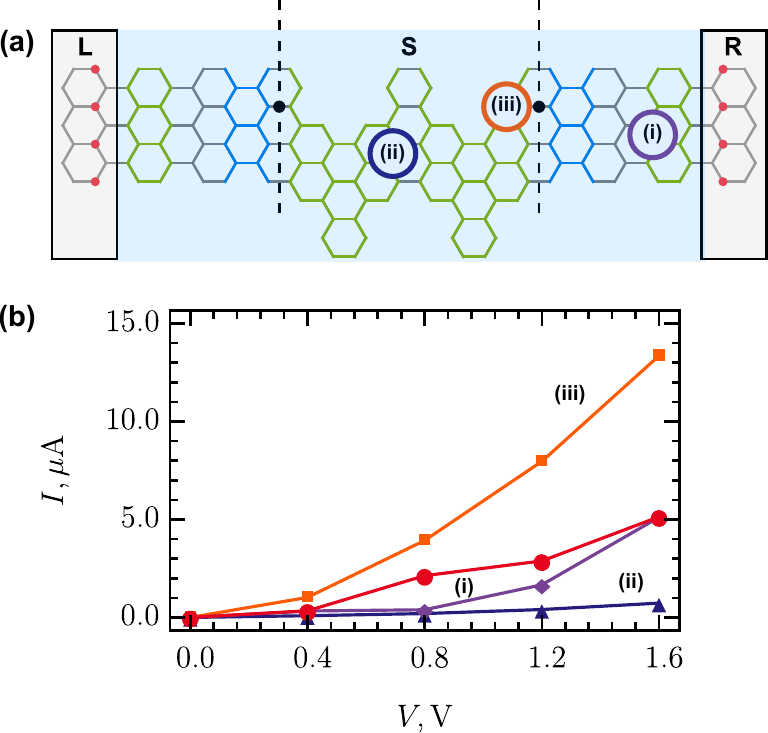}
  \caption{Quantum transport readout of gas molecules. (a) The scheme of the quantum transport sensor. Notations are the same as in Table~\ref{tbl:TypesOfGNRJunctions} and Figure~\ref{fig:DJDStructure}. The red dots denote nitrogen atoms. `L', `S', and `R' are the left lead, scattering region (whole light blue area), and right lead, respectively. (b) The current-voltage characteristics of the sensor before and after \ce{NO2} adsorption: (i) at AGNR($7$), (ii) at A60$(2,2,4,4)$, and (iii) the topological junction between A60$(2,2,4,4)$ and AGNR($7$) unit cells. The red [disks] are data for pristine sensor, while violet [rhombuses], dark blue [triangles] and orange [squares] are data for the sites that are marked with the same colors in panel (a).}
  \label{fig:NO2adsorptionReadOut}
\end{figure}

{\bf Sensor read-out with coherent transport.} The constructed symmetric DJ has a favorable configuration for connecting leads to and passing a coherent current through it. This current must be sensitive to the electronic properties of the scattering region represented by two junctions between topological and trivial nanoribbon unit cells. The changing current shall be a signature of the physical adsorption happening at the junction. The geometry of the leads is crucial here. In order to facilitate electron injection into the semiconducting structure, we use a combination of topological and chemical engineering. Specifically, we terminate the scattering region with topologically nontrivial AGNR($7$) unit cell; this engineering is equivalent to that Fig.~4 of Ref.~\citenum{Lee2018} or the one in the recent experiment on Janus GNRs~\cite{Song2025}. In addition, we $n$-dope the left and right leads with N-atoms as shown in Figure~\ref{fig:NO2adsorptionReadOut}a to make them conducting; see Supporting Information Note~2 and Figures~S1-S2. The latter may be a bit challenging step from a technological point of view, though atomically precise doping of GNRs has been demonstrated~\cite{Cai2014,Friedrich2020,Blackwell2021,Wen2023,Jacobse2025} and N-C interfaces have already been studied experimentally~\cite{Chen2017d}.

The quantum transport through the lead-DJ-lead system, where leads are semi-infinite, is explored using DFT and non-equilibrium Green's function methods implemented in \emph{NanoDCAL} software~\cite{Taylor2001a,Taylor2001b}, see Ref.~\citenum{SM-nl-2025-001492} and Supporting Information Note~1. 
Working in a low-bias regime and accounting for well-matched lead-DJ interface with negligible relaxation and charge transfer, the non-self-consistent transport modeling is sufficient~\cite{Ke2004}.

In Figure~\ref{fig:NO2adsorptionReadOut}b, the current-voltage ($I$-$V$) characteristics for the DJ-based sensor are presented within a reasonable range of voltages from $0$ to $1.6$~V. The reverse bias must have similar effect because of the mirror symmetry of the DJ, while the negative bias for the leads is not recommended due to the $n$-doping of the leads and the band gap below the Fermi level. The modeled sensor demonstrates measurable currents of about tens of $\mu$A's. Due to not very long the AGNR($7$) and A60$(2,2,4,4)$ fragments and the lead engineering, the scattering region passes enough current. Longer segments of AGNR($7$) show lower conductance in our quantum transport simulations. The adsorption of \ce{NO2} at the favorable junction site results in a twofold increase in current. Simultaneously, the \ce{NO2} adsorption at energetically less favorable sites of AGNR($7$) and A60$(2,2,4,4)$ blocks the current through the sensor. The presented detection mechanism differs from that of the previously proposed junction-free AGNR-based sensors~\cite{Tamersit2019}. In our case, the adsorption of the molecule at the junction increases the current, whereas in other schemes the current is usually blocked. 

Finally, we estimate the recovery time to obtain critical insights into \ce{NO2} desorption dynamics of the constructed sensor and to reveal its effectiveness and long-term applicability~\cite{SM-nl-2025-001492}. The recovery time ($\tau$) is calculated using the transition state theory Arrhenius-type equation~\cite{Arrhenius1889,Parey2022}:
\begin{equation}
  \tau = \nu^{-1} \exp\left(-\frac{E_{\mathrm{ac}}}{k_{\mathrm{B}}T}\right)\, ,
  \label{eqn:RecoveryTime}
\end{equation}
where $\nu$ is the attempt frequency, $E_{\mathrm{a}}$ is the activation energy, $k_{\mathrm{B}}$ is Boltzmann's constant in eV/K and $T$ is the temperature in K. The value of the attempt frequency is taken to be $10^{12}$~s$^{-1}$ as suggested for carbon nanotube-based \ce{NO2} sensor in Ref.~\citenum{Peng2004}. 
At the room temperature $T=298$~K, the values of $\tau$ for adsorption sites (i), (ii) and (iii) are $3.68\times 10^{-8}$, $1.19 \times 10^{-7}$, and $0.15$~s, respectively. The recovery time for \ce{NO2} adsorption at the topological junction site (iii) is significantly longer than at the AGNR or A60 sites which means the sensor could be used as is, i.e. without capping these areas with protective layers. Concurrently, $0.15$~s is more than ten fold shorter than experimental $2$~s recovery times recently reported for \ce{In2O3}-nanoparticle-based \ce{NO2} sensor~\cite{Jin2023}. This suggests that our sensor could support a high frequency of measurements and rapid data collection.

{\bf Summary and Conclusions.} We report a systematic topological classification for an A60 class of graphene nanoribbons, which can be used in a mirror symmetric double junction configuration as active elements of a gas sensor that is read out by quantum transport measurements. This envisages a niche for sensor applications of topological materials, in addition to the quantum computing perspective suggested in the literature. In view of the recent advance in an experimental study of topological end states in germanene nanoribbons~\cite{Klaassen2025}, we infer that our calculations can also be extended to inorganic chemistry; or even metamaterials~\cite{Plotnik2014,SerraGarcia2018,Peri2019,Xia2023}.

\begin{acknowledgement}
The authors thank T.~Giovannini for useful discussions and help with some data visualization, and M.~Pizzochero for useful comments. 
This work is supported by the National Natural Science Foundation of China (No. 12274361, No. 12474276). This work is also supported by Ongoing Research Funding Program (ORF-2025-468) King Saud University, Riyadh, Saudi Arabia.
R.B.P. acknowledges the support from the Institute of Physics, University of the Philippines – Los Ba\~{n}os. V.A.S. was partly supported by HORIZON EUROPE MSCA-2021-PF-01 (Project No. 101065500, TeraExc). 
\end{acknowledgement}

\begin{suppinfo}
Details of our electronic structure and transport calculations, results on electronic and topological properties of the N-dopped sensor leads modeled at the tight-binding and density functional theory levels:
\href{https://doi.org/10.5281/zenodo.15672059}{https://doi.org/10.5281/zenodo.15672059}
\end{suppinfo}

\bibliography{library}

\providecommand{\latin}[1]{#1}
\makeatletter
\providecommand{\doi}
  {\begingroup\let\do\@makeother\dospecials
  \catcode`\{=1 \catcode`\}=2 \doi@aux}
\providecommand{\doi@aux}[1]{\endgroup\texttt{#1}}
\makeatother
\providecommand*\mcitethebibliography{\thebibliography}
\csname @ifundefined\endcsname{endmcitethebibliography}  {\let\endmcitethebibliography\endthebibliography}{}
\begin{mcitethebibliography}{79}
\providecommand*\natexlab[1]{#1}
\providecommand*\mciteSetBstSublistMode[1]{}
\providecommand*\mciteSetBstMaxWidthForm[2]{}
\providecommand*\mciteBstWouldAddEndPuncttrue
  {\def\EndOfBibitem{\unskip.}}
\providecommand*\mciteBstWouldAddEndPunctfalse
  {\let\EndOfBibitem\relax}
\providecommand*\mciteSetBstMidEndSepPunct[3]{}
\providecommand*\mciteSetBstSublistLabelBeginEnd[3]{}
\providecommand*\EndOfBibitem{}
\mciteSetBstSublistMode{f}
\mciteSetBstMaxWidthForm{subitem}{(\alph{mcitesubitemcount})}
\mciteSetBstSublistLabelBeginEnd
  {\mcitemaxwidthsubitemform\space}
  {\relax}
  {\relax}

\bibitem[Hasan and Kane(2010)Hasan, and Kane]{Hasan2010}
Hasan,~M.~Z.; Kane,~C.~L. {C}olloquium: {T}opological insulators. \emph{Rev. Mod. Phys.} \textbf{2010}, \emph{82}, 3045--3067\relax
\mciteBstWouldAddEndPuncttrue
\mciteSetBstMidEndSepPunct{\mcitedefaultmidpunct}
{\mcitedefaultendpunct}{\mcitedefaultseppunct}\relax
\EndOfBibitem
\bibitem[Ren \latin{et~al.}(2016)Ren, Qiao, and Niu]{Ren2016}
Ren,~Y.; Qiao,~Z.; Niu,~Q. {T}opological phases in two-dimensional materials: a review. \emph{Rep. Prog. Phys} \textbf{2016}, \emph{79}, 066501\relax
\mciteBstWouldAddEndPuncttrue
\mciteSetBstMidEndSepPunct{\mcitedefaultmidpunct}
{\mcitedefaultendpunct}{\mcitedefaultseppunct}\relax
\EndOfBibitem
\bibitem[Cayssol and Fuchs(2021)Cayssol, and Fuchs]{Cayssol2021}
Cayssol,~J.; Fuchs,~J.~N. {T}opological and geometrical aspects of band theory. \emph{J. Phys. Mater.} \textbf{2021}, \emph{4}, 034007\relax
\mciteBstWouldAddEndPuncttrue
\mciteSetBstMidEndSepPunct{\mcitedefaultmidpunct}
{\mcitedefaultendpunct}{\mcitedefaultseppunct}\relax
\EndOfBibitem
\bibitem[Weber \latin{et~al.}(2024)Weber, Fuhrer, Sheng, Yang, Thomale, Shamim, Molenkamp, Cobden, Pesin, Zandvliet, Bampoulis, Claessen, Menges, Gooth, Felser, Shekhar, Tadich, Zhao, Edmonds, Jia, Bieniek, Väyrynen, Culcer, Muralidharan, and Nadeem]{Weber2024}
Weber,~B. \latin{et~al.}  2024 {R}oadmap on {2D} Topological Insulators. \emph{J. Phys. Mater.} \textbf{2024}, \emph{7}, 022501\relax
\mciteBstWouldAddEndPuncttrue
\mciteSetBstMidEndSepPunct{\mcitedefaultmidpunct}
{\mcitedefaultendpunct}{\mcitedefaultseppunct}\relax
\EndOfBibitem
\bibitem[Qi and Zhang(2011)Qi, and Zhang]{Qi2011a}
Qi,~X.-L.~L.; Zhang,~S.-C.~C. {T}opological insulators and superconductors. \emph{Rev. Mod. Phys.} \textbf{2011}, \emph{83}, 1057--1110\relax
\mciteBstWouldAddEndPuncttrue
\mciteSetBstMidEndSepPunct{\mcitedefaultmidpunct}
{\mcitedefaultendpunct}{\mcitedefaultseppunct}\relax
\EndOfBibitem
\bibitem[Ando(2013)]{Ando2013}
Ando,~Y. Topological insulator materials. \emph{J. Phys. Soc. Jpn.} \textbf{2013}, \emph{82}, 102001\relax
\mciteBstWouldAddEndPuncttrue
\mciteSetBstMidEndSepPunct{\mcitedefaultmidpunct}
{\mcitedefaultendpunct}{\mcitedefaultseppunct}\relax
\EndOfBibitem
\bibitem[Wehling \latin{et~al.}(2014)Wehling, Black-Schaffer, and Balatsky]{Wehling2014a}
Wehling,~T.; Black-Schaffer,~A.; Balatsky,~A. {D}irac materials. \emph{Adv. Phys.} \textbf{2014}, \emph{63}, 1--76\relax
\mciteBstWouldAddEndPuncttrue
\mciteSetBstMidEndSepPunct{\mcitedefaultmidpunct}
{\mcitedefaultendpunct}{\mcitedefaultseppunct}\relax
\EndOfBibitem
\bibitem[Kane and Mele(2005)Kane, and Mele]{Kane2005a}
Kane,~C.~L.; Mele,~E.~J. {Q}uantum spin {H}all effect in graphene. \emph{Phys. Rev. Lett.} \textbf{2005}, \emph{95}, 226801\relax
\mciteBstWouldAddEndPuncttrue
\mciteSetBstMidEndSepPunct{\mcitedefaultmidpunct}
{\mcitedefaultendpunct}{\mcitedefaultseppunct}\relax
\EndOfBibitem
\bibitem[Kane and Mele(2005)Kane, and Mele]{Kane2005}
Kane,~C.~L.; Mele,~E.~J. {Z$_2$} topological order and the quantum spin {H}all effect. \emph{Phys. Rev. Lett.} \textbf{2005}, \emph{95}, 146802\relax
\mciteBstWouldAddEndPuncttrue
\mciteSetBstMidEndSepPunct{\mcitedefaultmidpunct}
{\mcitedefaultendpunct}{\mcitedefaultseppunct}\relax
\EndOfBibitem
\bibitem[Cao \latin{et~al.}(2017)Cao, Zhao, and Louie]{Cao2017}
Cao,~T.; Zhao,~F.; Louie,~S.~G. {T}opological phases in graphene nanoribbons: {J}unction States, spin centers, and quantum spin shains. \emph{Phys. Rev. Lett.} \textbf{2017}, \emph{119}, 076401\relax
\mciteBstWouldAddEndPuncttrue
\mciteSetBstMidEndSepPunct{\mcitedefaultmidpunct}
{\mcitedefaultendpunct}{\mcitedefaultseppunct}\relax
\EndOfBibitem
\bibitem[Lee \latin{et~al.}(2018)Lee, Zhao, Cao, Ihm, and Louie]{Lee2018}
Lee,~Y.-l.; Zhao,~F.; Cao,~T.; Ihm,~J.; Louie,~S.~G. {T}opological phases in cove-edged and chevron graphene nanoribbons: {G}eometric structures, $Z_2$ invariants, and junction states. \emph{Nano Lett.} \textbf{2018}, \emph{18}, 7247--7253\relax
\mciteBstWouldAddEndPuncttrue
\mciteSetBstMidEndSepPunct{\mcitedefaultmidpunct}
{\mcitedefaultendpunct}{\mcitedefaultseppunct}\relax
\EndOfBibitem
\bibitem[Lin and Chou(2018)Lin, and Chou]{Lin2018}
Lin,~K.-S.; Chou,~M.-Y. {T}opological properties of gapped graphene nanoribbons with spatial symmetries. \emph{Nano Lett.} \textbf{2018}, \emph{18}, 7254--7260\relax
\mciteBstWouldAddEndPuncttrue
\mciteSetBstMidEndSepPunct{\mcitedefaultmidpunct}
{\mcitedefaultendpunct}{\mcitedefaultseppunct}\relax
\EndOfBibitem
\bibitem[Ortiz \latin{et~al.}(2018)Ortiz, García-Martínez, Lado, and Fernández-Rossier]{Ortiz2018}
Ortiz,~R.; García-Martínez,~N.~A.; Lado,~J.~L.; Fernández-Rossier,~J. {E}lectrical spin manipulation in graphene nanostructures. \emph{Phys. Rev. B} \textbf{2018}, \emph{97}, 195425\relax
\mciteBstWouldAddEndPuncttrue
\mciteSetBstMidEndSepPunct{\mcitedefaultmidpunct}
{\mcitedefaultendpunct}{\mcitedefaultseppunct}\relax
\EndOfBibitem
\bibitem[Rizzo \latin{et~al.}(2018)Rizzo, Veber, Cao, Bronner, Chen, Zhao, Rodriguez, Louie, Crommie, and Fischer]{Rizzo2018}
Rizzo,~D.~J.; Veber,~G.; Cao,~T.; Bronner,~C.; Chen,~T.; Zhao,~F.; Rodriguez,~H.; Louie,~S.~G.; Crommie,~M.~F.; Fischer,~F.~R. {T}opological band engineering of graphene nanoribbons. \emph{Nature} \textbf{2018}, \emph{560}, 204--208\relax
\mciteBstWouldAddEndPuncttrue
\mciteSetBstMidEndSepPunct{\mcitedefaultmidpunct}
{\mcitedefaultendpunct}{\mcitedefaultseppunct}\relax
\EndOfBibitem
\bibitem[Gr{\"o}ning \latin{et~al.}(2018)Gr{\"o}ning, Wang, Yao, Pignedoli, Borin~Barin, Daniels, Cupo, Meunier, Feng, Narita, Müllen, Ruffieux, and Fasel]{Groning2018}
Gr{\"o}ning,~O.; Wang,~S.; Yao,~X.; Pignedoli,~C.~A.; Borin~Barin,~G.; Daniels,~C.; Cupo,~A.; Meunier,~V.; Feng,~X.; Narita,~A.; Müllen,~K.; Ruffieux,~P.; Fasel,~R. {E}ngineering of robust topological quantum phases in graphene nanoribbons. \emph{Nature} \textbf{2018}, \emph{560}, 209--213\relax
\mciteBstWouldAddEndPuncttrue
\mciteSetBstMidEndSepPunct{\mcitedefaultmidpunct}
{\mcitedefaultendpunct}{\mcitedefaultseppunct}\relax
\EndOfBibitem
\bibitem[Joost \latin{et~al.}(2019)Joost, Jauho, and Bonitz]{Joost2019}
Joost,~J.~P.; Jauho,~A.~P.; Bonitz,~M. {C}orrelated topological states in graphene nanoribbon heterostructures. \emph{Nano Lett.} \textbf{2019}, \emph{19}, 9045--9050\relax
\mciteBstWouldAddEndPuncttrue
\mciteSetBstMidEndSepPunct{\mcitedefaultmidpunct}
{\mcitedefaultendpunct}{\mcitedefaultseppunct}\relax
\EndOfBibitem
\bibitem[Jiang and Louie(2021)Jiang, and Louie]{Jiang2021}
Jiang,~J.; Louie,~S.~G. Topology classification using chiral symmetry and spin correlations in graphene nanoribbons. \emph{Nano Lett.} \textbf{2021}, \emph{21}, 197--202\relax
\mciteBstWouldAddEndPuncttrue
\mciteSetBstMidEndSepPunct{\mcitedefaultmidpunct}
{\mcitedefaultendpunct}{\mcitedefaultseppunct}\relax
\EndOfBibitem
\bibitem[Arnold \latin{et~al.}(2022)Arnold, Liu, Kuc, and Heine]{Arnold2022}
Arnold,~F.~M.; Liu,~T.-J.; Kuc,~A.; Heine,~T. {S}tructure-imposed electronic topology in cove-edged graphene nanoribbons. \emph{Phys. Rev. Lett.} \textbf{2022}, \emph{129}, 216401\relax
\mciteBstWouldAddEndPuncttrue
\mciteSetBstMidEndSepPunct{\mcitedefaultmidpunct}
{\mcitedefaultendpunct}{\mcitedefaultseppunct}\relax
\EndOfBibitem
\bibitem[Kuo(2023)]{Kuo2023}
Kuo,~D. M.~T. {E}ffects of coulomb blockade on the charge transport through the topological states of finite armchair graphene nanoribbons and heterostructures. \emph{Nanomaterials} \textbf{2023}, \emph{13}, 1757\relax
\mciteBstWouldAddEndPuncttrue
\mciteSetBstMidEndSepPunct{\mcitedefaultmidpunct}
{\mcitedefaultendpunct}{\mcitedefaultseppunct}\relax
\EndOfBibitem
\bibitem[Perkins and Ferreira(2024)Perkins, and Ferreira]{Perkins2024}
Perkins,~D. T.~S.; Ferreira,~A. {U}ltrafast all-electrical universal nanoqubits. \emph{Phys. Rev. B} \textbf{2024}, \emph{109}, L041411\relax
\mciteBstWouldAddEndPuncttrue
\mciteSetBstMidEndSepPunct{\mcitedefaultmidpunct}
{\mcitedefaultendpunct}{\mcitedefaultseppunct}\relax
\EndOfBibitem
\bibitem[Wang \latin{et~al.}(2024)Wang, Yin, Tang, Du, Liang, Wang, Deng, Tan, Zhang, Ma, Tan, and Wang]{Wang2024}
Wang,~Z.; Yin,~R.; Tang,~Z.; Du,~H.; Liang,~Y.; Wang,~X.; Deng,~Q.-S.; Tan,~Y.-Z.; Zhang,~Y.; Ma,~C.; Tan,~S.; Wang,~B. {T}opologically localized vibronic excitations in second-layer graphene nanoribbons. \emph{Phys. Rev. Lett.} \textbf{2024}, \emph{133}, 036401\relax
\mciteBstWouldAddEndPuncttrue
\mciteSetBstMidEndSepPunct{\mcitedefaultmidpunct}
{\mcitedefaultendpunct}{\mcitedefaultseppunct}\relax
\EndOfBibitem
\bibitem[Zhao \latin{et~al.}(2024)Zhao, Catarina, Zhang, Henriques, Yang, Ma, Feng, Gröning, Ruffieux, Fern\'andez-Rossier, and Fasel]{Zhao2024}
Zhao,~C.; Catarina,~G.; Zhang,~J.-J.; Henriques,~J.~a. C.~G.; Yang,~L.; Ma,~J.; Feng,~X.; Gröning,~O.; Ruffieux,~P.; Fern\'andez-Rossier,~J.; Fasel,~R. {T}unable topological phases in nanographene-based spin-1/2 alternating-exchange {H}eisenberg chains. \emph{Nat. Nanotech.} \textbf{2024}, \emph{19}, 1789--1795\relax
\mciteBstWouldAddEndPuncttrue
\mciteSetBstMidEndSepPunct{\mcitedefaultmidpunct}
{\mcitedefaultendpunct}{\mcitedefaultseppunct}\relax
\EndOfBibitem
\bibitem[Yano \latin{et~al.}(2020)Yano, Mitoma, Ito, and Itami]{Yano2020}
Yano,~Y.; Mitoma,~N.; Ito,~H.; Itami,~K. {A} quest for structurally uniform graphene nanoribbons: {S}ynthesis, properties, and applications. \emph{J. Org. Chem.} \textbf{2020}, \emph{85}, 4--33\relax
\mciteBstWouldAddEndPuncttrue
\mciteSetBstMidEndSepPunct{\mcitedefaultmidpunct}
{\mcitedefaultendpunct}{\mcitedefaultseppunct}\relax
\EndOfBibitem
\bibitem[Chen \latin{et~al.}(2020)Chen, Narita, and M{\"u}llen]{Chen2020}
Chen,~Z.; Narita,~A.; M{\"u}llen,~K. {G}raphene nanoribbons: {On}-Surface synthesis and integration into electronic devices. \emph{Adv. Mater.} \textbf{2020}, \emph{32}, 2001893\relax
\mciteBstWouldAddEndPuncttrue
\mciteSetBstMidEndSepPunct{\mcitedefaultmidpunct}
{\mcitedefaultendpunct}{\mcitedefaultseppunct}\relax
\EndOfBibitem
\bibitem[Simon(1983)]{Simon1983}
Simon,~B. {H}olonomy, the quantum adiabatic theorem, and {B}erry's phase. \emph{Phys. Rev. Lett.} \textbf{1983}, \emph{51}, 2167--2170\relax
\mciteBstWouldAddEndPuncttrue
\mciteSetBstMidEndSepPunct{\mcitedefaultmidpunct}
{\mcitedefaultendpunct}{\mcitedefaultseppunct}\relax
\EndOfBibitem
\bibitem[Niu \latin{et~al.}(1985)Niu, Thouless, and Wu]{Niu1985}
Niu,~Q.; Thouless,~D.~J.; Wu,~Y.-S. {Q}uantized {H}all conductance as a topological invariant. \emph{Phys. Rev. B} \textbf{1985}, \emph{31}, 3372--3377\relax
\mciteBstWouldAddEndPuncttrue
\mciteSetBstMidEndSepPunct{\mcitedefaultmidpunct}
{\mcitedefaultendpunct}{\mcitedefaultseppunct}\relax
\EndOfBibitem
\bibitem[Haldane(1988)]{Haldane1988}
Haldane,~F. D.~M. Model for a quantum {H}all effect without landau levels: Condensed-matter realization of the "parity anomaly". \emph{Phys. Rev. Lett.} \textbf{1988}, \emph{61}, 2015--2018\relax
\mciteBstWouldAddEndPuncttrue
\mciteSetBstMidEndSepPunct{\mcitedefaultmidpunct}
{\mcitedefaultendpunct}{\mcitedefaultseppunct}\relax
\EndOfBibitem
\bibitem[Wang \latin{et~al.}(2016)Wang, Talirz, Pignedoli, Feng, M{\"u}llen, Fasel, and Ruffieux]{Wang2016a}
Wang,~S.; Talirz,~L.; Pignedoli,~C.~A.; Feng,~X.; M{\"u}llen,~K.; Fasel,~R.; Ruffieux,~P. {G}iant edge state splitting at atomically precise graphene zigzag edges. \emph{Nat. Commun.} \textbf{2016}, \emph{7}, 11507\relax
\mciteBstWouldAddEndPuncttrue
\mciteSetBstMidEndSepPunct{\mcitedefaultmidpunct}
{\mcitedefaultendpunct}{\mcitedefaultseppunct}\relax
\EndOfBibitem
\bibitem[Gao \latin{et~al.}(2016)Gao, Gao, Yang, Dai, Zhou, Zhang, and Lieber]{Gao2016}
Gao,~N.; Gao,~T.; Yang,~X.; Dai,~X.; Zhou,~W.; Zhang,~A.; Lieber,~C.~M. {S}pecific detection of biomolecules in physiological solutions using graphene transistor biosensors. \emph{Proc. Natl. Acad. Sci. U.S.A.} \textbf{2016}, \emph{113}, 14633--14638\relax
\mciteBstWouldAddEndPuncttrue
\mciteSetBstMidEndSepPunct{\mcitedefaultmidpunct}
{\mcitedefaultendpunct}{\mcitedefaultseppunct}\relax
\EndOfBibitem
\bibitem[Adamu \latin{et~al.}(2021)Adamu, Chen, and Chu]{Adamu2021}
Adamu,~B.~I.; Chen,~P.; Chu,~W. Role of nanostructuring of sensing materials in performance of electrical gas sensors by combining with extra strategies. \emph{Nano Ex.} \textbf{2021}, \emph{2}, 042003\relax
\mciteBstWouldAddEndPuncttrue
\mciteSetBstMidEndSepPunct{\mcitedefaultmidpunct}
{\mcitedefaultendpunct}{\mcitedefaultseppunct}\relax
\EndOfBibitem
\bibitem[Johnson \latin{et~al.}(2021)Johnson, Sabu, Swamy, Anto, Gangadharappa, and Pramod]{Johnson2021}
Johnson,~A.~P.; Sabu,~C.; Swamy,~N.~K.; Anto,~A.; Gangadharappa,~H.; Pramod,~K. {G}raphene nanoribbon: {An} emerging and efficient flat molecular platform for advanced biosensing. \emph{Biosens. Bioelectron} \textbf{2021}, \emph{184}, 113245\relax
\mciteBstWouldAddEndPuncttrue
\mciteSetBstMidEndSepPunct{\mcitedefaultmidpunct}
{\mcitedefaultendpunct}{\mcitedefaultseppunct}\relax
\EndOfBibitem
\bibitem[Abdelsalam \latin{et~al.}(2019)Abdelsalam, Saroka, and Younis]{Abdelsalam2019b}
Abdelsalam,~H.; Saroka,~V.~A.; Younis,~W.~O. {E}dge functionalization of finite graphene nanoribbon superlattices. \emph{Superlattices Microstruct.} \textbf{2019}, \emph{129}, 54--61\relax
\mciteBstWouldAddEndPuncttrue
\mciteSetBstMidEndSepPunct{\mcitedefaultmidpunct}
{\mcitedefaultendpunct}{\mcitedefaultseppunct}\relax
\EndOfBibitem
\bibitem[{\v C}er{\c n}evi{\v c}s \latin{et~al.}(2020){\v C}er{\c n}evi{\v c}s, Pizzochero, and Yazyev]{Cernevics2020}
{\v C}er{\c n}evi{\v c}s,~K.; Pizzochero,~M.; Yazyev,~O.~V. {E}ven–odd conductance effect in graphene nanoribbons induced by edge functionalization with aromatic molecules: {B}asis for novel chemosensors. \emph{Eur. Phys. J. Plus} \textbf{2020}, \emph{135}, 681\relax
\mciteBstWouldAddEndPuncttrue
\mciteSetBstMidEndSepPunct{\mcitedefaultmidpunct}
{\mcitedefaultendpunct}{\mcitedefaultseppunct}\relax
\EndOfBibitem
\bibitem[Abdelsalam \latin{et~al.}(2021)Abdelsalam, Saroka, Teleb, Ali, Osman, and Zhang]{Abdelsalam2021}
Abdelsalam,~H.; Saroka,~V.; Teleb,~N.; Ali,~M.; Osman,~W.; Zhang,~Q. {E}lectronic and adsorption properties of extended chevron and cove-edged graphene nanoribbons. \emph{Physica E Low Dimens.} \textbf{2021}, \emph{126}, 114438\relax
\mciteBstWouldAddEndPuncttrue
\mciteSetBstMidEndSepPunct{\mcitedefaultmidpunct}
{\mcitedefaultendpunct}{\mcitedefaultseppunct}\relax
\EndOfBibitem
\bibitem[Sakr \latin{et~al.}(2024)Sakr, Abdelsalam, Teleb, Abd-Elkader, Saroka, and Zhang]{Sakr2024}
Sakr,~M.~A.; Abdelsalam,~H.; Teleb,~N.~H.; Abd-Elkader,~O.~H.; Saroka,~V.~A.; Zhang,~Q. {I}nvestigating adsorption characteristics and electronic properties of {C}lar’s goblet and beyond. \emph{Chem. Phys. Lett.} \textbf{2024}, \emph{849}, 141428\relax
\mciteBstWouldAddEndPuncttrue
\mciteSetBstMidEndSepPunct{\mcitedefaultmidpunct}
{\mcitedefaultendpunct}{\mcitedefaultseppunct}\relax
\EndOfBibitem
\bibitem[Mehdi~Pour \latin{et~al.}(2017)Mehdi~Pour, Lashkov, Radocea, Liu, Sun, Lipatov, Korlacki, Shekhirev, Aluru, Lyding, Sysoev, and Sinitskii]{MehdiPour2017}
Mehdi~Pour,~M.; Lashkov,~A.; Radocea,~A.; Liu,~X.; Sun,~T.; Lipatov,~A.; Korlacki,~R.~A.; Shekhirev,~M.; Aluru,~N.~R.; Lyding,~J.~W.; Sysoev,~V.; Sinitskii,~A. {L}aterally extended atomically precise graphene nanoribbons with improved electrical conductivity for efficient gas sensing. \emph{Nat. Commun.} \textbf{2017}, \emph{8}, 820\relax
\mciteBstWouldAddEndPuncttrue
\mciteSetBstMidEndSepPunct{\mcitedefaultmidpunct}
{\mcitedefaultendpunct}{\mcitedefaultseppunct}\relax
\EndOfBibitem
\bibitem[Cho \latin{et~al.}(2018)Cho, Cho, Chong, Koh, Kim, Kim, and Jung]{Cho2018}
Cho,~K.~M.; Cho,~S.-Y.; Chong,~S.; Koh,~H.-J.; Kim,~D.~W.; Kim,~J.; Jung,~H.-T. {E}dge-functionalized graphene nanoribbon chemical sensor: {C}omparison with carbon nanotube and graphene. \emph{{ACS} Appl. Mater. Interfaces} \textbf{2018}, \emph{10}, 42905--42914\relax
\mciteBstWouldAddEndPuncttrue
\mciteSetBstMidEndSepPunct{\mcitedefaultmidpunct}
{\mcitedefaultendpunct}{\mcitedefaultseppunct}\relax
\EndOfBibitem
\bibitem[Saroka \latin{et~al.}(2014)Saroka, Batrakov, and Chernozatonskii]{Saroka2014a}
Saroka,~V.~A.; Batrakov,~K.~G.; Chernozatonskii,~L.~A. {E}dge-modified zigzag-shaped graphene nanoribbons: Structure and electronic properties. \emph{Phys. Solid State} \textbf{2014}, \emph{56}, 2135--2145\relax
\mciteBstWouldAddEndPuncttrue
\mciteSetBstMidEndSepPunct{\mcitedefaultmidpunct}
{\mcitedefaultendpunct}{\mcitedefaultseppunct}\relax
\EndOfBibitem
\bibitem[Saroka \latin{et~al.}(2015)Saroka, Batrakov, Demin, and Chernozatonskii]{Saroka2015}
Saroka,~V.~A.; Batrakov,~K.~G.; Demin,~V.~A.; Chernozatonskii,~L.~A. {B}and gaps in jagged and straight graphene nanoribbons tunable by an external electric field. \emph{J. Phys.: Condens. Matter} \textbf{2015}, \emph{27}, 145305\relax
\mciteBstWouldAddEndPuncttrue
\mciteSetBstMidEndSepPunct{\mcitedefaultmidpunct}
{\mcitedefaultendpunct}{\mcitedefaultseppunct}\relax
\EndOfBibitem
\bibitem[Saroka and Batrakov(2015)Saroka, and Batrakov]{Saroka2015a}
Saroka,~V.~A.; Batrakov,~K.~G. In \emph{Physics, Chemistry and Applications of Nanostructures}; Borisenko,~V.~E., Gaponenko,~S.~V., Gurin,~V.~S., Kam,~C.~H., Eds.; World Scientific: Singapore, 2015; pp 240--243\relax
\mciteBstWouldAddEndPuncttrue
\mciteSetBstMidEndSepPunct{\mcitedefaultmidpunct}
{\mcitedefaultendpunct}{\mcitedefaultseppunct}\relax
\EndOfBibitem
\bibitem[Saroka and Batrakov(2016)Saroka, and Batrakov]{Saroka2016a}
Saroka,~V.~A.; Batrakov,~K.~G. {Z}igzag-shaped Superlattices on the basis of graphene nanoribbons: {S}tructure and electronic properties. \emph{Russ. Phys. J.} \textbf{2016}, \emph{59}, 633--639\relax
\mciteBstWouldAddEndPuncttrue
\mciteSetBstMidEndSepPunct{\mcitedefaultmidpunct}
{\mcitedefaultendpunct}{\mcitedefaultseppunct}\relax
\EndOfBibitem
\bibitem[Kudin \latin{et~al.}(2007)Kudin, Car, and Resta]{Kudin2007}
Kudin,~K.~N.; Car,~R.; Resta,~R. {B}erry phase approach to longitudinal dipole moments of infinite chains in electronic-structure methods with local basis sets. \emph{J. Chem. Phys.} \textbf{2007}, \emph{126}, 234101\relax
\mciteBstWouldAddEndPuncttrue
\mciteSetBstMidEndSepPunct{\mcitedefaultmidpunct}
{\mcitedefaultendpunct}{\mcitedefaultseppunct}\relax
\EndOfBibitem
\bibitem[Rhim \latin{et~al.}(2017)Rhim, Behrends, and Bardarson]{Rhim2017}
Rhim,~J.-W.; Behrends,~J.; Bardarson,~J.~H. {B}ulk-boundary correspondence from the intercellular {Z}ak phase. \emph{Phys. Rev. B} \textbf{2017}, \emph{95}, 035421\relax
\mciteBstWouldAddEndPuncttrue
\mciteSetBstMidEndSepPunct{\mcitedefaultmidpunct}
{\mcitedefaultendpunct}{\mcitedefaultseppunct}\relax
\EndOfBibitem
\bibitem[Last \latin{et~al.}(1994)Last, Sun, and Witschi]{Last1994}
Last,~J.~A.; Sun,~W.~M.; Witschi,~H. Ozone, {NO}, and {NO}$_2$: oxidant air pollutants and more. \emph{Environ. Health Perspect.} \textbf{1994}, \emph{102}, 179--184\relax
\mciteBstWouldAddEndPuncttrue
\mciteSetBstMidEndSepPunct{\mcitedefaultmidpunct}
{\mcitedefaultendpunct}{\mcitedefaultseppunct}\relax
\EndOfBibitem
\bibitem[Huangfu and Atkinson(2020)Huangfu, and Atkinson]{Huangfu2020}
Huangfu,~P.; Atkinson,~R. Long-term exposure to {NO}$_2$ and O$_3$ and all-cause and respiratory mortality: A systematic review and meta-analysis. \emph{Environ. Int.} \textbf{2020}, \emph{144}, 105998\relax
\mciteBstWouldAddEndPuncttrue
\mciteSetBstMidEndSepPunct{\mcitedefaultmidpunct}
{\mcitedefaultendpunct}{\mcitedefaultseppunct}\relax
\EndOfBibitem
\bibitem[Li \latin{et~al.}(2022)Li, Zeng, and Li]{Li2022}
Li,~Q.; Zeng,~W.; Li,~Y. Metal oxide gas sensors for detecting {NO}$_2$ in industrial exhaust gas: Recent developments. \emph{Sens. Actuators B: Chem.} \textbf{2022}, \emph{359}, 131579\relax
\mciteBstWouldAddEndPuncttrue
\mciteSetBstMidEndSepPunct{\mcitedefaultmidpunct}
{\mcitedefaultendpunct}{\mcitedefaultseppunct}\relax
\EndOfBibitem
\bibitem[Sichau \latin{et~al.}(2019)Sichau, Prada, Anlauf, Lyon, Bosnjak, Tiemann, and Blick]{Sichau2019}
Sichau,~J.; Prada,~M.; Anlauf,~T.; Lyon,~T.~J.; Bosnjak,~B.; Tiemann,~L.; Blick,~R.~H. {R}esonance microwave measurements of an intrinsic spin-orbit coupling gap in graphene: {A} possible indication of a topological state. \emph{Phys. Rev. Lett.} \textbf{2019}, \emph{122}, 046403\relax
\mciteBstWouldAddEndPuncttrue
\mciteSetBstMidEndSepPunct{\mcitedefaultmidpunct}
{\mcitedefaultendpunct}{\mcitedefaultseppunct}\relax
\EndOfBibitem
\bibitem[Partoens and Peeters(2006)Partoens, and Peeters]{Partoens2006}
Partoens,~B.; Peeters,~F.~M. {F}rom graphene to graphite: Electronic structure around the K point. \emph{Phys. Rev. B} \textbf{2006}, \emph{74}, 075404\relax
\mciteBstWouldAddEndPuncttrue
\mciteSetBstMidEndSepPunct{\mcitedefaultmidpunct}
{\mcitedefaultendpunct}{\mcitedefaultseppunct}\relax
\EndOfBibitem
\bibitem[Saroka \latin{et~al.}(2018)Saroka, Abdelsalam, Demin, Grassano, Kuten, Pushkarchuk, and Pulci]{Saroka2018a}
Saroka,~V.~A.; Abdelsalam,~H.; Demin,~V.~A.; Grassano,~D.; Kuten,~S.~A.; Pushkarchuk,~A.; Pulci,~O. {A}bsorption in finite-length chevron-type graphene nanoribbons. \emph{Semiconductors} \textbf{2018}, \emph{52}, 1890--1893\relax
\mciteBstWouldAddEndPuncttrue
\mciteSetBstMidEndSepPunct{\mcitedefaultmidpunct}
{\mcitedefaultendpunct}{\mcitedefaultseppunct}\relax
\EndOfBibitem
\bibitem[Payod \latin{et~al.}(2020)Payod, Grassano, Santos, Levshov, Pulci, and Saroka]{Payod2020}
Payod,~R.~B.; Grassano,~D.; Santos,~G. N.~C.; Levshov,~D.~I.; Pulci,~O.; Saroka,~V.~A. {2N+4}-rule and an atlas of bulk optical resonances of zigzag graphene nanoribbons. \emph{Nat. Commun.} \textbf{2020}, \emph{11}, 82\relax
\mciteBstWouldAddEndPuncttrue
\mciteSetBstMidEndSepPunct{\mcitedefaultmidpunct}
{\mcitedefaultendpunct}{\mcitedefaultseppunct}\relax
\EndOfBibitem
\bibitem[Saroka(2025)]{TBpackSaroka}
Saroka,~V.~A. {TB}pack (version 0.5.2 and higher). 2025; \url{https://github.com/vasilsaroka/TBpack}, Accessed: 2025-04-25\relax
\mciteBstWouldAddEndPuncttrue
\mciteSetBstMidEndSepPunct{\mcitedefaultmidpunct}
{\mcitedefaultendpunct}{\mcitedefaultseppunct}\relax
\EndOfBibitem
\bibitem[SM-()]{SM-nl-2025-001492}
See Supplementary Materials at https://doi.org/10.5281/zenodo.15209274 that include TB and DFT calculation raw data and Mathematica code to reproduce the reported results\relax
\mciteBstWouldAddEndPuncttrue
\mciteSetBstMidEndSepPunct{\mcitedefaultmidpunct}
{\mcitedefaultendpunct}{\mcitedefaultseppunct}\relax
\EndOfBibitem
\bibitem[Tepliakov \latin{et~al.}(2023)Tepliakov, Lischner, Kaxiras, Mostofi, and Pizzochero]{Tepliakov2023}
Tepliakov,~N.~V.; Lischner,~J.; Kaxiras,~E.; Mostofi,~A.~A.; Pizzochero,~M. {U}nveiling and manipulating hidden symmetries in graphene nanoribbons. \emph{Phys. Rev. Lett.} \textbf{2023}, \emph{130}, 9045--9050\relax
\mciteBstWouldAddEndPuncttrue
\mciteSetBstMidEndSepPunct{\mcitedefaultmidpunct}
{\mcitedefaultendpunct}{\mcitedefaultseppunct}\relax
\EndOfBibitem
\bibitem[Cai \latin{et~al.}(2010)Cai, Ruffieux, Jaafar, Bieri, Braun, Blankenburg, Muoth, Seitsonen, Saleh, Feng, Müllen, and Fasel]{Cai2010}
Cai,~J.; Ruffieux,~P.; Jaafar,~R.; Bieri,~M.; Braun,~T.; Blankenburg,~S.; Muoth,~M.; Seitsonen,~A.~P.; Saleh,~M.; Feng,~X.; Müllen,~K.; Fasel,~R. {A}tomically precise bottom-up fabrication of graphene nanoribbons. \emph{Nature} \textbf{2010}, \emph{466}, 470--473\relax
\mciteBstWouldAddEndPuncttrue
\mciteSetBstMidEndSepPunct{\mcitedefaultmidpunct}
{\mcitedefaultendpunct}{\mcitedefaultseppunct}\relax
\EndOfBibitem
\bibitem[Son \latin{et~al.}(2006)Son, Cohen, and Louie]{Son2006a}
Son,~Y.-W.; Cohen,~M.~L.; Louie,~S.~G. {E}nergy gaps in graphene nanoribbons. \emph{Phys. Rev. Lett.} \textbf{2006}, \emph{97}, 216803\relax
\mciteBstWouldAddEndPuncttrue
\mciteSetBstMidEndSepPunct{\mcitedefaultmidpunct}
{\mcitedefaultendpunct}{\mcitedefaultseppunct}\relax
\EndOfBibitem
\bibitem[Frisch \latin{et~al.}(2016)Frisch, Trucks, Schlegel, Scuseria, Robb, Cheeseman, Scalmani, Barone, Petersson, Nakatsuji, Li, Caricato, Marenich, Bloino, Janesko, Gomperts, Mennucci, Hratchian, Ortiz, Izmaylov, Sonnenberg, Williams-Young, Ding, Lipparini, Egidi, Goings, Peng, Petrone, Henderson, Ranasinghe, Zakrzewski, Gao, Rega, Zheng, Liang, Hada, Ehara, Toyota, Fukuda, Hasegawa, Ishida, Nakajima, Honda, Kitao, Nakai, Vreven, Throssell, Montgomery, Peralta, Ogliaro, Bearpark, Heyd, Brothers, Kudin, Staroverov, Keith, Kobayashi, Normand, Raghavachari, Rendell, Burant, Iyengar, Tomasi, Cossi, Millam, Klene, Adamo, Cammi, Ochterski, Martin, Morokuma, Farkas, Foresman, and Fox]{g16}
Frisch,~M.~J. \latin{et~al.}  Gaussian~16 {R}evision {C}.01. 2016; Gaussian Inc. Wallingford CT\relax
\mciteBstWouldAddEndPuncttrue
\mciteSetBstMidEndSepPunct{\mcitedefaultmidpunct}
{\mcitedefaultendpunct}{\mcitedefaultseppunct}\relax
\EndOfBibitem
\bibitem[Lu and Chen(2012)Lu, and Chen]{Lu2012}
Lu,~T.; Chen,~F. {M}ultiwfn: {A} multifunctional wavefunction analyzer. \emph{J. Comput. Chem.} \textbf{2012}, \emph{33}, 580--592\relax
\mciteBstWouldAddEndPuncttrue
\mciteSetBstMidEndSepPunct{\mcitedefaultmidpunct}
{\mcitedefaultendpunct}{\mcitedefaultseppunct}\relax
\EndOfBibitem
\bibitem[Collier \latin{et~al.}(2017)Collier, Saroka, and Portnoi]{Collier2017a}
Collier,~T.~P.; Saroka,~V.~A.; Portnoi,~M.~E. {T}uning terahertz transitions in a double-gated quantum ring. \emph{Phys. Rev. B} \textbf{2017}, \emph{96}, 235430\relax
\mciteBstWouldAddEndPuncttrue
\mciteSetBstMidEndSepPunct{\mcitedefaultmidpunct}
{\mcitedefaultendpunct}{\mcitedefaultseppunct}\relax
\EndOfBibitem
\bibitem[Song \latin{et~al.}(2025)Song, Teng, Tang, Xu, He, Ruan, Kojima, Hu, Giessibl, Sakaguchi, Louie, and Lu]{Song2025}
Song,~S.; Teng,~Y.; Tang,~W.; Xu,~Z.; He,~Y.; Ruan,~J.; Kojima,~T.; Hu,~W.; Giessibl,~F.~J.; Sakaguchi,~H.; Louie,~S.~G.; Lu,~J. {J}anus graphene nanoribbons with localized states on a single zigzag edge. \emph{Nature} \textbf{2025}, \emph{637}, 580--586\relax
\mciteBstWouldAddEndPuncttrue
\mciteSetBstMidEndSepPunct{\mcitedefaultmidpunct}
{\mcitedefaultendpunct}{\mcitedefaultseppunct}\relax
\EndOfBibitem
\bibitem[Cai \latin{et~al.}(2014)Cai, Pignedoli, Talirz, Ruffieux, Söde, Liang, Meunier, Berger, Li, Feng, M{\"u}llen, and Fasel]{Cai2014}
Cai,~J.; Pignedoli,~C.~A.; Talirz,~L.; Ruffieux,~P.; Söde,~H.; Liang,~L.; Meunier,~V.; Berger,~R.; Li,~R.; Feng,~X.; M{\"u}llen,~K.; Fasel,~R. {G}raphene nanoribbon heterojunctions. \emph{Nat. Nanotechnol.} \textbf{2014}, \emph{9}, 896--900\relax
\mciteBstWouldAddEndPuncttrue
\mciteSetBstMidEndSepPunct{\mcitedefaultmidpunct}
{\mcitedefaultendpunct}{\mcitedefaultseppunct}\relax
\EndOfBibitem
\bibitem[Friedrich \latin{et~al.}(2020)Friedrich, Brandimarte, Li, Saito, Yamaguchi, Pozo, Pe{\~n}a, Frederiksen, Garcia-Lekue, S{\'a}nchez-Portal, and Pascual]{Friedrich2020}
Friedrich,~N.; Brandimarte,~P.; Li,~J.; Saito,~S.; Yamaguchi,~S.; Pozo,~I.; Pe{\~n}a,~D.; Frederiksen,~T.; Garcia-Lekue,~A.; S{\'a}nchez-Portal,~D.; Pascual,~J.~I. {M}agnetism of topological boundary states induced by boron substitution in graphene nanoribbons. \emph{Phys. Rev. Lett.} \textbf{2020}, \emph{125}, 146801\relax
\mciteBstWouldAddEndPuncttrue
\mciteSetBstMidEndSepPunct{\mcitedefaultmidpunct}
{\mcitedefaultendpunct}{\mcitedefaultseppunct}\relax
\EndOfBibitem
\bibitem[Blackwell \latin{et~al.}(2021)Blackwell, Zhao, Brooks, Zhu, Piskun, Wang, Delgado, Lee, Louie, and Fischer]{Blackwell2021}
Blackwell,~R.~E.; Zhao,~F.; Brooks,~E.; Zhu,~J.; Piskun,~I.; Wang,~S.; Delgado,~A.; Lee,~Y.-L.; Louie,~S.~G.; Fischer,~F.~R. {Spin} splitting of dopant edge state in magnetic zigzag graphene nanoribbons. \emph{Nature} \textbf{2021}, \emph{600}, 647--652\relax
\mciteBstWouldAddEndPuncttrue
\mciteSetBstMidEndSepPunct{\mcitedefaultmidpunct}
{\mcitedefaultendpunct}{\mcitedefaultseppunct}\relax
\EndOfBibitem
\bibitem[Wen \latin{et~al.}(2023)Wen, Jacobse, Jiang, Wang, Louie, Crommie, and Fischer]{Wen2023}
Wen,~E. C.~H.; Jacobse,~P.~H.; Jiang,~J.; Wang,~Z.; Louie,~S.~G.; Crommie,~M.~F.; Fischer,~F.~R. {F}ermi-Level engineering of nitrogen core-doped armchair graphene nanoribbons. \emph{J. Am. Chem. Soc.} \textbf{2023}, \emph{145}, 19338--19346\relax
\mciteBstWouldAddEndPuncttrue
\mciteSetBstMidEndSepPunct{\mcitedefaultmidpunct}
{\mcitedefaultendpunct}{\mcitedefaultseppunct}\relax
\EndOfBibitem
\bibitem[Jacobse \latin{et~al.}(2025)Jacobse, Pizzochero, Wen, Barin, Li, Mutlu, Müllen, Kaxiras, Crommie, and Fischer]{Jacobse2025}
Jacobse,~P.~H.; Pizzochero,~M.; Wen,~E. C.~H.; Barin,~G.~B.; Li,~X.; Mutlu,~Z.; Müllen,~K.; Kaxiras,~E.; Crommie,~M.~F.; Fischer,~F.~R. {C}oupling of nondegenerate topological modes in nitrogen core-doped graphene nanoribbons. \emph{{ACS} Nano} \textbf{2025}, \emph{19}, 13029--13036\relax
\mciteBstWouldAddEndPuncttrue
\mciteSetBstMidEndSepPunct{\mcitedefaultmidpunct}
{\mcitedefaultendpunct}{\mcitedefaultseppunct}\relax
\EndOfBibitem
\bibitem[Chen \latin{et~al.}(2017)Chen, He, Wang, Wang, Tang, Cong, Xie, Li, Xia, Li, Wu, Zhang, Deng, Yu, Xie, and Jiang]{Chen2017d}
Chen,~L. \latin{et~al.}  {O}riented graphene nanoribbons embedded in hexagonal boron nitride trenches. \emph{Nat. Commun.} \textbf{2017}, \emph{8}, 14703\relax
\mciteBstWouldAddEndPuncttrue
\mciteSetBstMidEndSepPunct{\mcitedefaultmidpunct}
{\mcitedefaultendpunct}{\mcitedefaultseppunct}\relax
\EndOfBibitem
\bibitem[Taylor \latin{et~al.}(2001)Taylor, Guo, and Wang]{Taylor2001a}
Taylor,~J.; Guo,~H.; Wang,~J. \textit{Ab initio} modeling of quantum transport properties of molecular electronic devices. \emph{Phys. Rev. B} \textbf{2001}, \emph{63}, 245407\relax
\mciteBstWouldAddEndPuncttrue
\mciteSetBstMidEndSepPunct{\mcitedefaultmidpunct}
{\mcitedefaultendpunct}{\mcitedefaultseppunct}\relax
\EndOfBibitem
\bibitem[Taylor \latin{et~al.}(2001)Taylor, Guo, and Wang]{Taylor2001b}
Taylor,~J.; Guo,~H.; Wang,~J. Ab initio modeling of open systems: Charge transfer, electron conduction, and molecular switching of a C$_{60}$ device. \emph{Phys. Rev. B} \textbf{2001}, \emph{63}, 121104\relax
\mciteBstWouldAddEndPuncttrue
\mciteSetBstMidEndSepPunct{\mcitedefaultmidpunct}
{\mcitedefaultendpunct}{\mcitedefaultseppunct}\relax
\EndOfBibitem
\bibitem[Ke \latin{et~al.}(2004)Ke, Baranger, and Yang]{Ke2004}
Ke,~S.-H.; Baranger,~H.~U.; Yang,~W. {E}lectron transport through molecules: {S}elf-consistent and non-self-consistent approaches. \emph{Phys. Rev. B} \textbf{2004}, \emph{70}, 085410\relax
\mciteBstWouldAddEndPuncttrue
\mciteSetBstMidEndSepPunct{\mcitedefaultmidpunct}
{\mcitedefaultendpunct}{\mcitedefaultseppunct}\relax
\EndOfBibitem
\bibitem[Tamersit(2019)]{Tamersit2019}
Tamersit,~K. {An} ultra-sensitive gas nanosensor based on asymmetric dual-gate graphene nanoribbon field-effect transistor: proposal and investigation. \emph{J. Comput. Electron.} \textbf{2019}, \emph{18}, 846--855\relax
\mciteBstWouldAddEndPuncttrue
\mciteSetBstMidEndSepPunct{\mcitedefaultmidpunct}
{\mcitedefaultendpunct}{\mcitedefaultseppunct}\relax
\EndOfBibitem
\bibitem[Arrhenius(1889)]{Arrhenius1889}
Arrhenius,~S. {\"U}ber die dissociationsw{\"a}rme und den einfluss der temperatur auf den dissociationsgrad der elektrolyte. \emph{Z. Phys. Chem.} \textbf{1889}, \emph{4U}, 96--116\relax
\mciteBstWouldAddEndPuncttrue
\mciteSetBstMidEndSepPunct{\mcitedefaultmidpunct}
{\mcitedefaultendpunct}{\mcitedefaultseppunct}\relax
\EndOfBibitem
\bibitem[Parey \latin{et~al.}(2022)Parey, Abraham, Gaur, and Thapa]{Parey2022}
Parey,~V.; Abraham,~B.~M.; Gaur,~N.~K.; Thapa,~R. {F}irst-principles study of two-dimensional B-doped carbon nanostructures for toxic phosgene gas detection. \emph{{ACS} Appl. Nano Mater.} \textbf{2022}, \emph{5}, 12737--12745\relax
\mciteBstWouldAddEndPuncttrue
\mciteSetBstMidEndSepPunct{\mcitedefaultmidpunct}
{\mcitedefaultendpunct}{\mcitedefaultseppunct}\relax
\EndOfBibitem
\bibitem[Peng \latin{et~al.}(2004)Peng, Cho, Qi, and Dai]{Peng2004}
Peng,~S.; Cho,~K.; Qi,~P.; Dai,~H. {Ab} initio study of {CNT} {NO}$_2$ gas sensor. \emph{Chem. Phys. Lett.} \textbf{2004}, \emph{387}, 271--276\relax
\mciteBstWouldAddEndPuncttrue
\mciteSetBstMidEndSepPunct{\mcitedefaultmidpunct}
{\mcitedefaultendpunct}{\mcitedefaultseppunct}\relax
\EndOfBibitem
\bibitem[Jin \latin{et~al.}(2023)Jin, Wang, Wu, Song, Yao, Liu, Zhao, Zeng, and Wang]{Jin2023}
Jin,~Z.; Wang,~C.; Wu,~L.; Song,~H.; Yao,~X.; Liu,~J.; Zhao,~J.; Zeng,~Z.; Wang,~F. {Fast} responding and recovering of {NO}$_2$ sensors based on Ni-doped In$_2$O$_3$ nanoparticles. \emph{Sens. Actuators B: Chem.} \textbf{2023}, \emph{377}, 133058\relax
\mciteBstWouldAddEndPuncttrue
\mciteSetBstMidEndSepPunct{\mcitedefaultmidpunct}
{\mcitedefaultendpunct}{\mcitedefaultseppunct}\relax
\EndOfBibitem
\bibitem[Klaassen \latin{et~al.}(2025)Klaassen, Eek, Rudenko, van’t Westende, Castenmiller, Zhang, De~Boeij, Van~Houselt, Ezawa, Zandvliet, Morais~Smith, and Bampoulis]{Klaassen2025}
Klaassen,~D.~J.; Eek,~L.; Rudenko,~A.~N.; van’t Westende,~E.~D.; Castenmiller,~C.; Zhang,~Z.; De~Boeij,~P.~L.; Van~Houselt,~A.; Ezawa,~M.; Zandvliet,~H. J.~W.; Morais~Smith,~C.; Bampoulis,~P. {R}ealization of a one-dimensional topological insulator in ultrathin germanene nanoribbons. \emph{Nat. Commun.} \textbf{2025}, \emph{16}, 2059\relax
\mciteBstWouldAddEndPuncttrue
\mciteSetBstMidEndSepPunct{\mcitedefaultmidpunct}
{\mcitedefaultendpunct}{\mcitedefaultseppunct}\relax
\EndOfBibitem
\bibitem[Plotnik \latin{et~al.}(2014)Plotnik, Rechtsman, Song, Heinrich, Zeuner, Nolte, Lumer, Malkova, Xu, Szameit, Chen, and Segev]{Plotnik2014}
Plotnik,~Y.; Rechtsman,~M.~C.; Song,~D.; Heinrich,~M.; Zeuner,~J.~M.; Nolte,~S.; Lumer,~Y.; Malkova,~N.; Xu,~J.; Szameit,~A.; Chen,~Z.; Segev,~M. Observation of unconventional edge states in ‘photonic graphene'. \emph{Nat. Mater.} \textbf{2014}, \emph{13}, 57--62\relax
\mciteBstWouldAddEndPuncttrue
\mciteSetBstMidEndSepPunct{\mcitedefaultmidpunct}
{\mcitedefaultendpunct}{\mcitedefaultseppunct}\relax
\EndOfBibitem
\bibitem[Serra-Garcia \latin{et~al.}(2018)Serra-Garcia, Peri, S{\"u}sstrunk, Bilal, Larsen, Villanueva, and Huber]{SerraGarcia2018}
Serra-Garcia,~M.; Peri,~V.; S{\"u}sstrunk,~R.; Bilal,~O.~R.; Larsen,~T.; Villanueva,~L.~G.; Huber,~S.~D. {O}bservation of a phononic quadrupole topological insulator. \emph{Nature} \textbf{2018}, \emph{555}, 342--345\relax
\mciteBstWouldAddEndPuncttrue
\mciteSetBstMidEndSepPunct{\mcitedefaultmidpunct}
{\mcitedefaultendpunct}{\mcitedefaultseppunct}\relax
\EndOfBibitem
\bibitem[Peri \latin{et~al.}(2019)Peri, Serra-Garcia, Ilan, and Huber]{Peri2019}
Peri,~V.; Serra-Garcia,~M.; Ilan,~R.; Huber,~S.~D. {A}xial-field-induced chiral channels in an acoustic {Weyl} system. \emph{Nat. Phys.} \textbf{2019}, \emph{15}, 357--361\relax
\mciteBstWouldAddEndPuncttrue
\mciteSetBstMidEndSepPunct{\mcitedefaultmidpunct}
{\mcitedefaultendpunct}{\mcitedefaultseppunct}\relax
\EndOfBibitem
\bibitem[Xia \latin{et~al.}(2023)Xia, Liang, Tang, Song, Xu, and Chen]{Xia2023}
Xia,~S.; Liang,~Y.; Tang,~L.; Song,~D.; Xu,~J.; Chen,~Z. {P}hotonic realization of a generic type of graphene edge states exhibiting topological flat band. \emph{Phys. Rev. Lett.} \textbf{2023}, \emph{131}, 013804\relax
\mciteBstWouldAddEndPuncttrue
\mciteSetBstMidEndSepPunct{\mcitedefaultmidpunct}
{\mcitedefaultendpunct}{\mcitedefaultseppunct}\relax
\EndOfBibitem
\end{mcitethebibliography}

\end{document}